\documentclass[preprint]{revtex4-1}
\usepackage{graphicx,subfigure,amsmath,bbm}
\usepackage{dcolumn}%
\usepackage{bm}%
\usepackage{amsmath}
\usepackage{amssymb}
\usepackage{multirow}
\usepackage{xcolor,soul}
\soulregister\cite7
\soulregister\ref7
\soulregister\pageref7
\usepackage{ulem}
\renewcommand{\hl}{}
\draft

\begin{document}

\author{Riccardo Capelli}
\author{Guido Tiana}
\email{guido.tiana@unimi.it}
\affiliation{Center for Complexity and Biosystems and Department of Physics, Universit\`a degli Studi di Milano and INFN, via Celoria 16, I-20133 Milano, Italy}

\author{Carlo Camilloni}
\email{carlo.camilloni@unimi.it}
\affiliation{Dipartimento di Bioscienze, Universit\`a degli Studi di Milano, via Celoria 26, I-20133 Milano, Italy}

\title{An implementation of the maximum--caliber principle by replica--averaged time--resolved restrained simulations}

\date{\today}

\begin{abstract}
Inferential methods can be used to integrate experimental informations and molecular simulations. The maximum entropy principle provides a framework for using equilibrium experimental data and it has been shown that replica--averaged  simulations, restrained using a static potential, are a practical and powerful implementation of such principle. Here we show that replica--averaged  simulations restrained using a time--dependent potential are equivalent to the principle of maximum caliber, the dynamic version of the principle of maximum entropy, and thus may allow to integrate time--resolved data in molecular dynamics simulations. We provide an analytical proof of the equivalence as well as a computational validation making use of simple models and synthetic data. Some limitations and possible solutions are also discussed.
\end{abstract}

\maketitle

\section{Introduction}

Molecular dynamics (MD) is a powerful sampling strategy that allow studying equilibrium as well as time--resolved properties of complex systems at atomistic resolution\cite{Dror:2012cs}. The predicting power of MD is related to both the quality of the force fields as well as to the extent of the sampling\cite{Zuckerman:2011dz}. 
Nowadays, the microsecond timescale is routinely accessible for systems of the order of 10 kDa, with the notable exception of Anton computers that allow performing simulations one to two order of magnitude longer\cite{Dror:2012cs}. 
\hl{When molecular events cannot be sampled by standard MD, the sampling can be enhanced by methods focused either on the recovery of the underlying free--energy\cite{Zuckerman:2011dz,Pietrucci:2017dy}, most notably Umbrella Sampling\cite{Torrie:1977wa}, or on the generation of reactive trajectories, like Markov-state models \cite{Husic:2018iz} and path-sampling methods\cite{Bolhuis:2002uk}}.
Modern force--fields can often reproduce quantitatively the equilibrium properties of small to medium--sized  proteins, even if the results are still often system and/or force--field dependent, in particular for disordered proteins\cite{Beauchamp:2012eq,LindorffLarsen:2012gl,Piana:2014jb,MartinGarcia:2015ji}. Force--fields robustness in reproducing kinetic properties is, instead, more questionable and poorly investigated\cite{Piana:2011dl,Vitalini:2015gc}.

In order to improve the accuracy of molecular simulations with respect to equilibrium properties in a system--specific way, hybrid methods based on the integration of experimental data in MD simulations have been introduced\cite{Torda:1989va,Vendruscolo:2003uk,LindorffLarsen:2005bi,Camilloni:2012je,Bonomi:2017dn}. These methods take into account the ensemble averaged nature of equilibrium experimental data by including additional energy terms to the force--field based on a forward model of the experimental observable and a bias that imposes the average agreement to the data either following the maximum entropy principle (pMaxEnt)\cite{Jaynes:1957ua,Pitera:2012ds,Roux:2013bp,Cavalli:2013jf,White:2014fd,Cesari:2016ix} or Bayesian statistics\cite{Hummer:2015ba,Bonomi:2016ip} and can be used to obtain results of comparable quality independently by the specific force--field\cite{Lohr:2017gc}. Hybrid approaches based on a statistical treatment of experimental data have been recently used also in combination with enhanced sampling methods\cite{Camilloni:2013di,Bonomi:2016ge}, ab--initio models\cite{White:2017ho}, coarse--grained models\cite{Hocky:2017fy} and Markov--state models\cite{Olsson:2017cn}. 

In principle, an inferential approach like the principle of maximum caliber (pMaxCal)\cite{Jaynes:1980hr}, that is the dynamic version of the principle of maximum entropy, could also be used to improve the quality of simulations in reproducing time--resolved properties. The pMaxCal was so far used to study basic aspects of non-equilibrium systems \cite{Stock:2008ii,Hazoglou:2015bs}, to model chemical reactions\cite{Presse:2011ja} and more recently do find collective variables for enhanced sampling techniques\cite{Tiwary:2016hw} and to reweigh the results of MD simulations\cite{Wan:2016fz,Zhou:2017bh} and of Markov State Models\cite{Dixit:2018hg} 
\hl{also out-of-equilibrium\cite{Dixit:2018iy}}. 
With respect to the MaxEnt\cite{Cesari:2018in} there is not yet an implementation that allows the direct integration of experimental data in MD simulations making use of a bias.

The pMaxCal states that the least--biased distribution $p(\gamma)$ of trajectories $\gamma$ generated by a stochastic process, like that associated with the dissipative dynamics of a biomolecule, is that obtained maximizing the path entropy (for an exhaustive review, see ref. \cite{Presse:2013dh,Dixit:2018by})
\begin{equation}
S[p(\gamma)]=-\sum_\gamma p(\gamma)\log p(\gamma).
\label{eq:maxcal}
\end{equation}

Similarly to what is done in equilibrium statistical mechanics, it is possible to use Lagrange multipliers to constrain the optimization of $S[p]$ in such a way that the average $\sum_\gamma p(\gamma)f(\gamma)$ of some conformational property $f$ of the system matches at each time any function of time (e.g. a function which reports the time course of some experimental data). The resulting distribution $p(\gamma)$, beside being in agreement with the experimental data, guarantees to minimize the amount of further, arbitrary information provided to the model.
 
In this work we present an implementation of the pMaxCal inspired by the replica-averaging implementation of the pMaxEnt\cite{Roux:2013bp,Cavalli:2013jf} that could allow to generate MD trajectories biased by time--resolved experimental data. 
\hl{The goal of such bias is not immediately that of generating more efficiently reactive trajectories, like in the case of path sampling methods, but that of improving the average agreement of an ensemble of MD simulations with an experimental time trace}. 
We first analytically showed the equivalence of the pMaxCal with replica--averaged time--resolved restrained simulations and then we used structure--based potentials and synthetic data to assess the reliability of replica-simulations in modulating time--resolved properties using multiple conformational parameters. We anticipate that one limit of the current approach is that one should be able to run MD of length comparable to that of the time--resolved observables of interest. Since time--resolved experimental observables report on processes often happening on longer time scales then those accessible by MD we use our simple--models to discuss the possibility of rescaling the time--scale of the guiding observable so to effectively rescale the time--scale of the 
\hl{ensemble of simulations}.

\section{Methods}

\subsection{Theoretical framework}

Our goal is to simulate the ensemble of trajectories that, initiate from a given state (a single conformation or an ensemble of conformations), follow the time course of a set of time--dependent experimental data and minimize the subjective bias introduced into the system, maximizing the associated caliber. We define $\{\gamma\}$ as the set of trajectories of the system, where the trajectories are regarded as discrete set of conformations $\gamma\equiv\{r_0,r_1,...,r_T\}$, as those usually generated in MD simulations. Kinetic experiments usually return time--resolved quantities that depend on the conformations visited along the trajectory. We define $f^{exp}_t$ the time--course of the quantity monitored in the available experiment, indexed by the discrete time $t$; this can be one-- or higher--dimensional. We assume to know the forward model associated with the experiment, that is the function $f(r_t)$ that maps a conformation $r_t$ visited along a trajectory into the ideal result that the experiment would give if applied to an ensemble of identical conformations $r_t$. Moreover, we assume to know the microscopic diffusion coefficient $D$ associated with the degrees of freedom of the system, for example obtaining it from specific experiments (like DOESY spectra from NMR experiments) or approximating it by Stokes' law.

In detail, given $\{\gamma\}$ our  set of stochastic trajectories of the $N$--particle system starting at point $r_0$, we are interested in the probability $p(\gamma)$. The principle of maximum caliber requires that  $p(r_0,r_1,...,r_T)$ maximizes
\begin{equation}
S[p]=-\sum_{\{\gamma\}}p(\{\gamma\})\log p(\{\gamma\})
\end{equation}
with the constraints
\begin{equation}
\sum_{\{\gamma\}}p(\{\gamma\})f(r_t)=f^{exp}_t
\label{eq:av}
\end{equation}
and
\begin{equation}
\frac{1}{2\Delta t}\sum_{\{\gamma\}}p(\{\gamma\})[r_{t+1}-r_t]^2=D
\label{eq:av2}
\end{equation}
at each discrete time $t$, and that $\sum p(\{\gamma\})=1$. One should note that any drift due to forces acting on the atoms scales as $\Delta t$ and does not contribute to Eq. (\ref{eq:av2}) in the limit of small $\Delta t$. The constrained maximization gives
\begin{equation}
p(\{\gamma\})=\frac{1}{Z_d}\exp\left[ -\sum_t \left(\nu_t [r_{t+1}-r_t]^2+\lambda_t f(r_t)\right) \right],
\label{eq:maxexp}
\end{equation}
where $Z_d$ is the normalization constant and $\nu_t$ is the set of Lagrange multipliers which implement the average of Eq. (\ref{eq:av2}) and $\lambda_t$ that implementing Eq. (\ref{eq:av}). In principle, $\lambda_t$ can be obtained by $d(\log Z_d)/d\lambda_t=f^{exp}_t$, but in practice this is hampered by the sum $Z_d$ over all possible paths.

It is useful to extend the expression found in Eq. (\ref{eq:maxexp}) in two ways. First, let us consider $n$ independent replicas of the system, each defined by trajectories $\{\gamma^\alpha\}=\{r_t^\alpha\}$ with $\alpha=1,...,n$ and $t=0,...,T$. The maximum-caliber probability distribution is then extended to
\begin{equation}
p(\{\gamma^\alpha\})=\frac{1}{Z_d}\exp\left[ -\sum_{t,\alpha} \left(\nu^\alpha_t [r_{t+1}^\alpha-r_t^\alpha]^2+\lambda^\alpha_t f(r_t^\alpha)\right) \right].
\label{eq:maxexp2}
\end{equation}
Moreover, one can require that 
\begin{equation}
\sum_{\{r_t^\alpha\}}p(\{\gamma^\alpha\}) \left[ \frac{1}{n}\sum_\alpha f(r_t^\alpha)-f^{exp}_t \right]^2 =\sigma_{nt}^2,
\end{equation}
that is that the standard error of the {\it average}  of $f$ over the replicas is some value $\sigma_n$. For sake of compactness, let's define
\begin{equation}
\xi_t\equiv \frac{1}{n}\sum_\beta f(r^\beta_t)-f^{exp}_t,
\end{equation}
implying that the experimental data are matched if $\xi_t=0$ for all $t$.
Applying the Lagrange--multipliers method also to this constrain, the maximum--caliber distribution becomes
\begin{equation}
p(\{\gamma^\alpha\})=\frac{1}{Z_d}\exp\left[ -\sum_{t,\alpha} \left(\nu^\alpha_t [r_{t+1}^\alpha-r_t^\alpha]^2+
\lambda^\alpha_t f(r_t^\alpha)  + \mu_{nt}^\alpha   \xi_t^2  \right) \right].
\label{eq:maxexp2}
\end{equation}
In the limit $n\to\infty$, $\sigma_{nt}\to 0$ for every $t$ because of the law of large numbers, and consequently one can set $\mu_{nt}^\alpha\to\infty$ for each $t$ and $\alpha$. In particular, $\sigma_n\sim n^{-1/2}$ and consequently $\mu^\alpha_{nt}\sim\log n$.

Similarly to the case of equilibrium simulations \cite{Roux:2013bp,Cavalli:2013jf}, we want to show that the maximum--caliber distribution of trajectories of Eq. (\ref{eq:maxexp}) is automatically sampled by replica--averaged MD simulations, with replicas (identified by greek letters) biased by a time--dependent potential

\begin{equation}
U(\{r^\alpha\},t)=\frac{nk}{2} \left( \frac{1}{n}\sum_\alpha f(r^\alpha)-f^{exp}_t  \right)^2,
\label{eq:bias}
\end{equation}
where $r^\alpha$ is the conformation of the system in the replica $\alpha$, $n$ is the number of replicas and $k$ is an harmonic constant. The associated stochastic process in the $(3N\times n)$--dimensional replica space can be regarded as a Markov chain 
\begin{equation}
p_{n}(\{r_t^\alpha\})=p_N(r_0^\alpha)w(r_0^\alpha\to r_1^\alpha)w(r_1^\alpha\to r_2^\alpha)...w(r_{T-1}^\alpha\to r_T^\alpha)
\end{equation}
which can be written according to the simplest form of the Onsager--Machlup function, corresponding to an over--damped stochastic dynamics discretized according to Ito prescription \cite{Adib:2008cp}
\begin{equation}
p_{n}(\{\gamma^\alpha\})=c\cdot \exp\left[ -\sum_{t\alpha}\frac{ \left(r_{t+1}^\alpha-r_t^\alpha+ k\Delta t\xi_t  \right)^2 }{2D'\Delta t}  \right],
\label{eq:om}
\end{equation}
recalling that by definition the initial point $r_0^\alpha$ is fixed for all replicas. Here the diffusion coefficient is $D'=T/\gamma'$, where $\gamma'$ is the friction coefficient chosen as an input of the simulation.
In the limit of large $k$ this can be approximated as
\begin{equation}
p_{n}(\{\gamma^\alpha\})=c\cdot \exp\left[ -\sum_{t\alpha}\frac{ \left( r_{t+1}^\alpha-r_t^\alpha  \right)^2 }{2D'\Delta t}  \right]\cdot \prod_{t}\delta\left( \xi_t   \right)
\label{eq:delta}
\end{equation}
because of the definition of Dirac's delta, that is  for any distribution $\varphi(\xi)$ and any $t$
\begin{equation}
c\int d\xi_t\;\exp\left[ -\sum_{\alpha}\frac{ \left(r_{t+1}^\alpha-r_t^\alpha+ k\Delta t\xi_t  \right)^2 }{2D'\Delta t}  \right] \varphi(\xi_t)= 
c\cdot \exp\left[ -\sum_{\alpha}\frac{ \left( r_{t+1}^\alpha-r_t^\alpha  \right)^2 }{2D'\Delta t}  \right]\cdot \varphi(0)
\end{equation}
in the limit $k\to\infty$.

Equation (\ref{eq:delta}) can be rewritten multiplying its r.h.s. by the exponential of a linear function of $\xi_t$, that is
which is equivalent to
\begin{equation}
p_{n}(\{\gamma^\alpha\})=c\cdot\exp\left[ -\sum_{t\alpha}\frac{ \left( r_{t+1}^\alpha-r_t^\alpha  \right)^2 }{2D'\Delta t} -\sum_t \gamma_t  \xi_t   \right]
\cdot\prod_{t}\delta\left( \xi_t   \right)
\label{eq:first}
\end{equation}
for any $\gamma_t$. In fact, for any distribution $\varphi(\xi)$ and any $t$
\begin{align}
&c\int d\xi_t\; \exp\left[ -\sum_{\alpha}\frac{ \left( r_{t+1}^\alpha-r_t^\alpha  \right)^2 }{2D'\Delta t}  \right]\delta\left( \xi_t   \right)\varphi(\xi_t)=\nonumber\\
&=c\int d\xi_t\;  \exp\left[ -\sum_{\alpha}\frac{ \left( r_{t+1}^\alpha-r_t^\alpha  \right)^2 }{2D'\Delta t} -\gamma_t \xi_t   \right]\delta\left( \xi_t  \right)\varphi(\xi_t),
\end{align}
meaning that Eq. (\ref{eq:delta}) is equivalent to Eq. (\ref{eq:first}).

Using the Gaussian representation of Dirac's delta $\delta(\xi_t)=\lim_{\kappa\to\infty}\exp(-\kappa\xi_t^2)$, Eq. (\ref{eq:first}) becomes
\begin{equation}
p_{n}(\{\gamma^\alpha\})=c\cdot \exp\left[ -\sum_{t\alpha}\frac{ \left( r_{t+1}^\alpha-r_t^\alpha  \right)^2 }{2D'\Delta t} -\sum_t \gamma_t \xi_t 
 - \sum_t \kappa_t \left( \xi_t  \right)^2\right]
\label{eq:second}
\end{equation}
in the limit $\kappa_t\to\infty$ for any $t$. Choosing $\gamma_t=\lambda_t$, remembering that both $\mu^\alpha_{nt}$ and $\kappa_t\to\infty$ for large $k$, then Eq. (\ref{eq:second}) is equivalent to the maximum--caliber distribution of Eq. (\ref{eq:maxexp2}).

However, there is a further difficulty involving the diffusion coefficient. If the experimental data are not taken into account, i.e.  $\lambda^\alpha_t=\mu^\alpha_{nt}=0$, then the partition function in Eq. (\ref{eq:maxexp2}) is a Gaussian integral and the condition $\partial\log Z_d/\partial \nu^\alpha_t=D$ defining the Lagrange multipliers gives $\nu_t^\alpha=1/D$ and thus $D=D'$. In this case, the diffusion coefficient used as an input to the replica simulation is the same required by the maximum--caliber principle.

On the other hand, if one accounts for the experimental data, then $\nu_t^\alpha \neq 1/D$ and the simulated diffusion of the particles becomes different from that required by the principle of maximum caliber. If the constraining effect of the experimental data is mild, one can expect that $\lambda_t^\alpha$ are small and the dynamical partition function in Eq. (\ref{eq:maxexp2}) can be approximated as
\begin{equation}
Z_d=\sum_{\{\gamma^\alpha\}} \exp\left[ -\sum_{t,\alpha} \nu^\alpha_t [r_{t+1}^\alpha-r_t^\alpha]^2  \right]
\left( 1-\sum_{t,\alpha} \lambda^\alpha_t f(r_t^\alpha) \right),
\label{eq:zapp}
\end{equation}
and consequently to the first order in $\lambda_t^\alpha$
\begin{equation}
D=\frac{1}{\nu_t^\alpha}-  \lambda^\alpha_t \frac{\partial}{\partial \nu_t^\alpha} \left\langle f(r_t^\alpha) \right\rangle_d,
\end{equation}
where $\langle \cdot \rangle_d$ is the unperturbed average over paths. Comparing this with Eq. (\ref{eq:second}) gives
\begin{equation}
D=D'  +  \lambda^\alpha_t (D')^2 \frac{\partial}{\partial D'} \left\langle f(r_t^\alpha) \right\rangle_d,
\end{equation}
suggesting that the actual diffusion coefficient is modified by the bias.

So, given the possibility to perform simulations on the same time scale of a time--resolved experiment, it is in theory possible to integrate the information of the  experimental time--course and generate trajectories in accord with the pMaxCal by means of replica--averaged time--resolved restrained simulations. 

\hl{Of notice, the theory in its present form is developed for the case of a uniform prior, nonetheless in the following we show that its implementation works also for the general case where a prior approximated Hamiltonian is available (e.g. a molecular mechanics force--field.}

\subsection{Validation strategy}

To test the validity of the replica--averaging time--resolved scheme on molecular models,  we performed some sand--box studies selecting some protein systems and defining for each of them two different structure--based G\=o potentials\cite{Whitford:2009ts}. One of the two ($U_{\text{ref}}$) is regarded as the reference potential that controls the dynamics of the system in our ideal experiment while the other ($U_{\text{approx}}$) is regarded as an approximated potential we known. The two potentials are chosen in such a way that the system displays markedly different kinetic properties when interacting with each of them, but similar equilibrium properties, this is what is somehow expected by current state-of-the-art force-fields. Structure based potentials allow us running a large number of simulations in a relatively short time making them perfectly suitable as a first step  towards a better understanding of the present time--resolved replica--averaging approach.

We performed multiple simulations with $U_{\text{ref}}$ that serve as reference for the tests. We also defined some conformational parameter $f^{\text{ref}}_t$ as our time--resolved synthetic observable that is obtained by averaging at each time step over the ensemble of simulation. Some of them (like the RMSD or the fraction of native contacts) are good approximations of the reaction coordinates of the system, while others (like the SAXS intensities) are closer to what one could measure in real experiments.

\begin{figure}[htp]
\includegraphics[width=\textwidth]{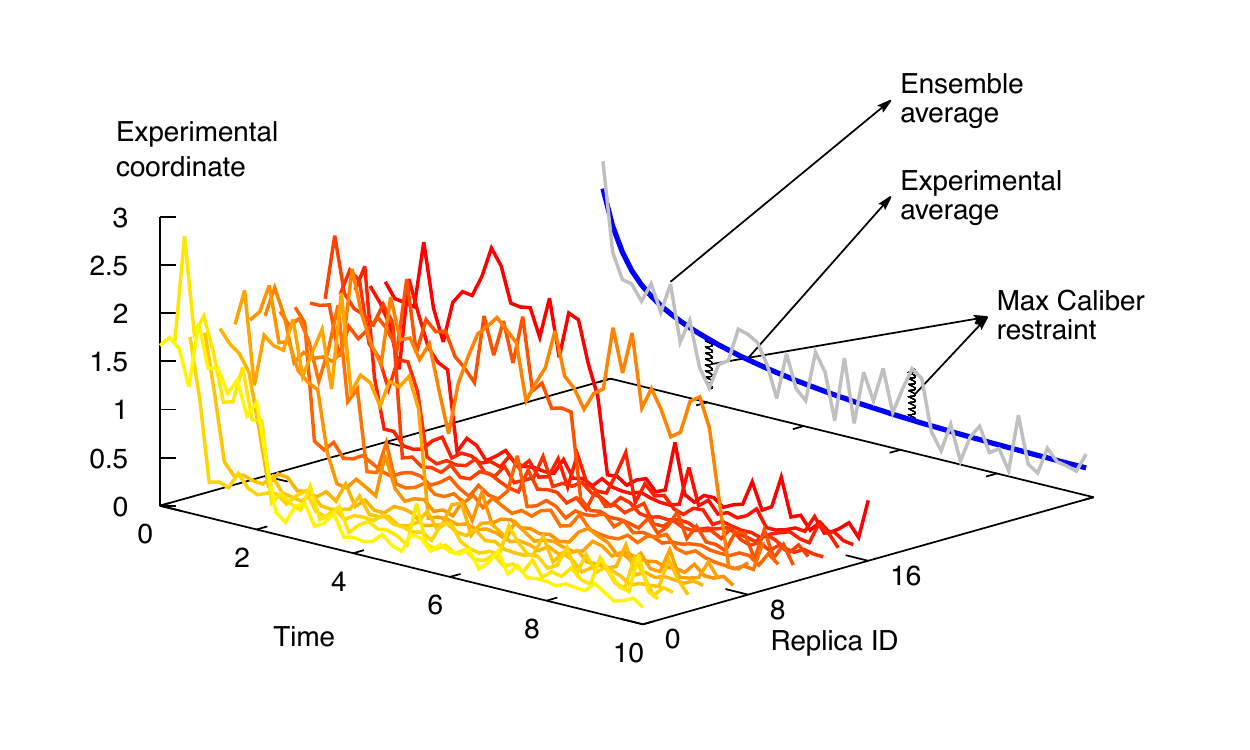}
\caption{A sketch of the MD simulations, where $n$ replicas of the system evolve in time coupled by Eq. (\protect\ref{eq:bias}). Lines colored in different hues of red and yellow represent the time evolution of the biasing variable in the various replicas. The grey line is the average of the biasing variable over the replicas. The biasing potential is an harmonic spring acting on this average, centred at the value of the experimental value (blue line) at the corresponding time.}
\label{3dplot_average}
\end{figure}

We applied the pMaxCal to the system interacting with the potential $U_{\text{approx}}$, performing MD simulations of $n$ replicas of the system biased by $f^{\text{ref}}_t$ through the potential described in Eq. (\ref{eq:bias}) (cf. Fig. \ref{3dplot_average}). The dynamics of the biasing variable averaged over the replicas, of its fluctuations over the replicas and of other variables weakly coupled to it are then compared with the reference dynamics.

\subsection{Computational Implementation}
\label{sect:comp}

MD simulations are performed with Gromacs 4.5.7\cite{Pronk:2013ef} coupled to Plumed 2\cite{Tribello:2014eb} using the ISDB module\cite{Bonomi:2017gca}. We implemented a  \texttt{CALIBER} bias into Plumed to apply the potential described in Eq. (\ref{eq:bias}). Simulations were performed with a Langevin integrator with $\gamma=1\text{ ps}^{\text{-1}}$ and a time--step of  0.1 fs.

We tested different quantities to bias the simulations, such as the root mean square deviation (RMSD) of the position of the C$_\alpha$ from those of the crystallographic conformation,  the fraction $Q$ of native contacts, defined as\cite{Best:2013hu} $Q(r) = \frac{1}{N} \sum_{i \neq j} \frac{1}{1 + \exp(\beta(r_{ij} - \lambda r_{ij}^0))} $, where $N$ is the total number of pairs in the potential, $r_{ij}\equiv|r_i-r_j|$ is the distance between the $i$-th and $j$-th atom, $r_{ij}^{0}$ is the distance between the two atoms in the crystallographic structure, $\beta=50\text{ nm}^{-1}$ and $\lambda=1.8$ are two switching parameters; and the theoretical SAXS intensities defined as $ I(q) = \sum_{i} \sum_{j \neq i} f_{i}(q) f_{j}(q) \frac{ \sin(q r_{ij}) }{qr_{ij}} $, where $q$ is the scattering vector, $f_{k}(q)$ is the atomic form factor of the $k$-th atom, and $r_{ij}$ is the distance between the $i$-th and the $j$-th atom.

The values of the harmonic constant $k$ were chosen to be as large as possible, compatibly with the time step of the simulation.

\section{Results}

\subsection{Modulation of the dynamics of a $\beta$--hairpin model}
\label{sect:hairpin}

\begin{figure}[htp]
\centering
\includegraphics[width=\textwidth]{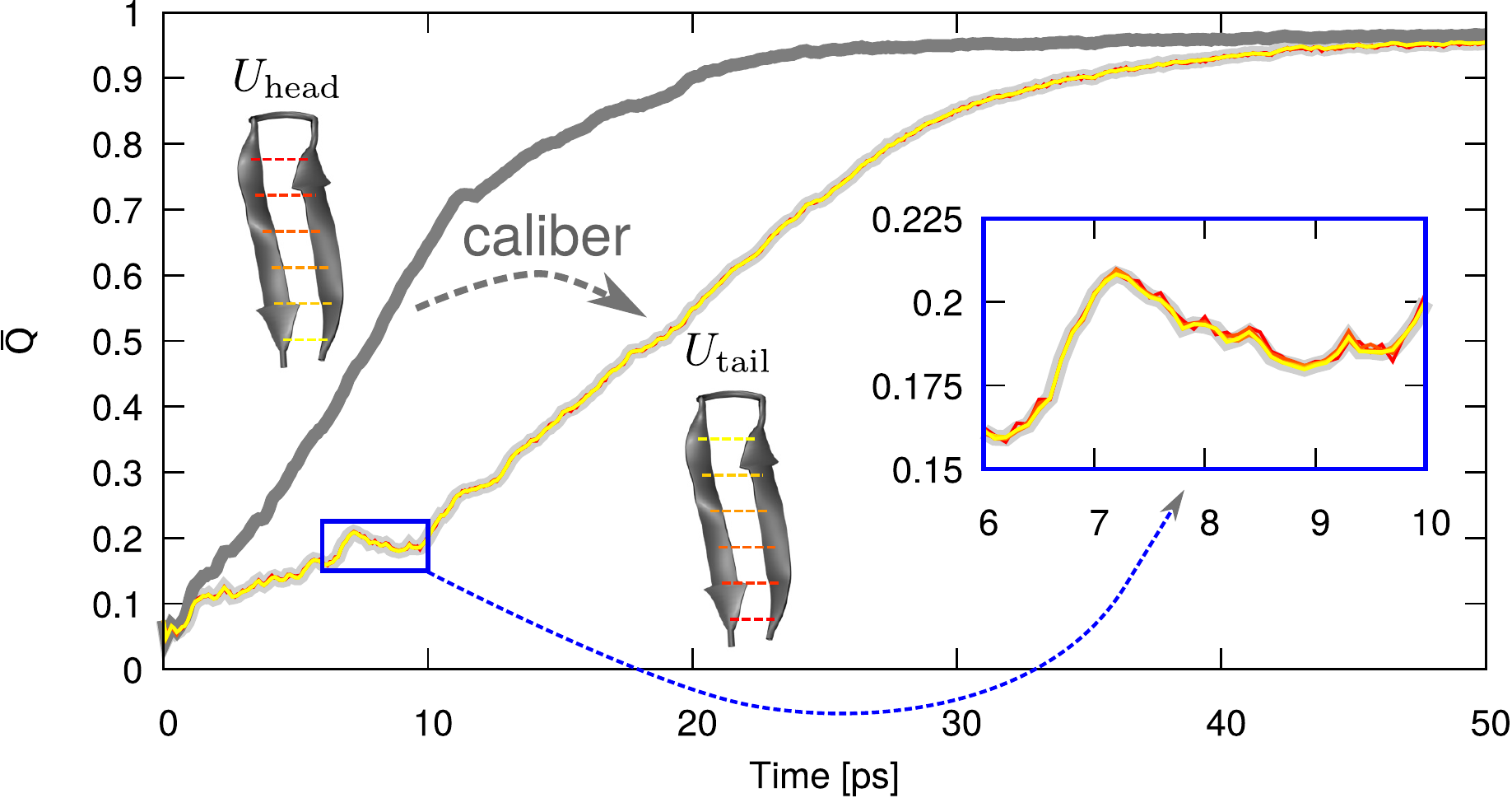}
\caption{\hl{MaxCal restraint over the time evolution of the average fraction of native contacts. A reference potential $U_{\text{tail}}$ is built assigning to the pairs of residues towards the turn of the hairpin weaker interactions than those towards the termini; the scaling factor of the G\=o interactions goes from 0.5 (yellow dashed lines) to 1.5 (red dashed lines). An approximated potential $U_{\text{head}}$ is built instead assigning to the pairs of residues towards the turn of the hairpin stronger interactions than those towards the termini; the scaling factor of the G\=o interactions goes from 1.5 (red dashed lines) to 0.5 (yellow dashed lines). The time evolution of the average fraction of native contacts $\overline{Q}$  is shown in light grey and dark grey for $U_{\text{tail}}$ and $U_{\text{head}}$, respectively. $\overline{Q}$ from $U_{\text{tail}}$ is used as the experimental observable to bias the approximated Hamiltonian $U_{\text{head}}$ by varying the number of replicas from 4 (red) to 128  (yellow), better visible in the inset.}}
\label{fig_q_tc}
\end{figure}

The first test to verify the ability of replica--averaged time--resolved simulations to modify the dynamics of a molecular system were carried out on an all--atom model of the second hairpin of protein G B1 domain (residues 41--65, pdb code 1PGB \cite{Gallagher:1994wt}) {\it in vacuo}. We built two different structure--based potentials\cite{Whitford:2009ts}, these potentials stabilize by definition a reference conformation. 
\hl{The potential $U_{\text{tail}}$ is obtained rescaling the interactions between the pairs of atoms of a factor which is proportional to the distance from the turn of the hairpin, from 0.5 for pairs close to the turn, to 1.5 for pairs close to the termini (see the hairpin schemes in Fig. \ref{fig_q_tc}). The potential $U_{\text{head}}$ is obtained inverting the scaling factors to strengthen by a factor 1.5 the interactions close to the turn and weaken by 1.5 those close to the termini, this induces a different folding dynamics while keeping comparable stability between the folded and the unfolded state (cf. the heat capacities displayed in Fig. S1). The dynamics of the hairpin interacting with both potentials was simulated starting from an unfolded conformation at $T=50$K (note that in a G\=o model energy units, and consequently temperature units, are arbitrary), generating 500 folding trajectories for each of them. In Fig. \ref{fig_q_tc} it displayed the average value ${\overline Q(t)}$ of the fraction of native contacts as a function of time, which result qualitatively different for the two systems (dark and light grey for $U_{\text{tail}}$ and $U_{\text{head}}$, respectively.}

The test consisted in biasing the system interacting with $U_{\text{head}}$  (regarded as $U_{\text{approx}}$) to display the dynamics of the system interacting with $U_{\text{tail}}$ (regarded as $U_{\text{ref}}$). For this purpose, we used the function ${\overline Q(t)}$ of the latter as reference data $f^{\text{ref}}(t)$, and simulated the dynamics of the hairpin with the potential $U_{\text{tail}}+V_{bias}$, varying the number of replicas from $n=4$ to $n=128$ and using a harmonic constant for $V_{bias}$ equal to $k=2.5\cdot10^{4} \cdot n$. The behavior of ${\overline Q(t)}$ for the resulting simulations is essentially indistinguishable from that of the simulations we wanted to target for any $n$, indicating that the two dynamics are identical at least when projected over the space defined by the biasing variable \hl{(cf. Fig. \ref{fig_q_tc})}. 

\hl{To check if not only the biased observable but also other observables are modified correctly upon the addition of the bias, we plotted the time evolution of the mean gyration radius and its standard deviation (cf. Fig. \ref{chi_q_tc_other}) as well as other unbiased observables (left panel of Fig. S2 of the Supplementary Materials). Also in this case, the biased curves match reasonably well the reference dynamics simulated with $U_{\text{tail}}$, quite independently on the number of replicas (cf. also the $\chi^2$ displayed in Figs. S3 and S4 of the Supplementary Materials).}

\hl{In addition to the average we also checked the effect on the fluctuations of the same observables.} In the lower panel of Fig. \ref{chi_q_tc_other},  we plotted the fluctuations of the gyration radius, defined as its standard deviation over the replicas as a function of time (cf. also the right panel of Fig. S2 for the standard deviation of other quantities). In spite of their noisy behavior, the bias is able to push the system interacting with  $U_{\text{head}}$ to display fluctuations similar to those of the system interacting with $U_{\text{tail}}$. Also for them there is not a clear behavior as a function of the number $n$ of replicas, except for the fact that $n=4$ gives an agreement that is much worse than for larger $n$ (see also Figs. S5--S6 in the Supplementary Materials). Finally, as a control, similar results are obtained by using $U_{\text{head}}$ as reference potential and biasing the system interacting with  $U_{\text{tail}}$ to follow its dynamics (see  Figs. S7--S15 in the Supplementary Materials).

\subsection{Modulation and rescaling of the dynamics of a simple protein model}
\label{sect:protG}

\begin{figure}[htp]
\centering
\includegraphics[width=0.9\textwidth]{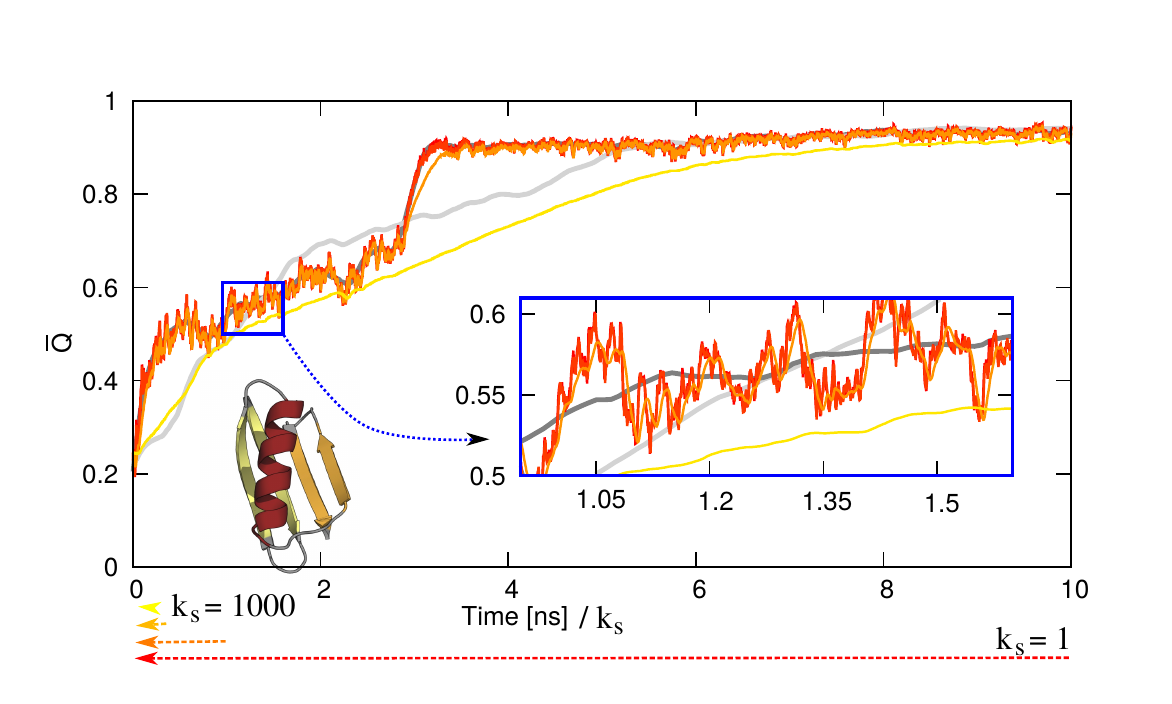}
\caption{Average fraction of native contacts $\overline{Q}$ as function of time for: the unbiased simulations of protein G interacting with $U_{\text{G\=o}}$ (dark grey); the unbiased simulations interacting with $U_{\alpha}$ (light grey); and for three biased simulations of the molecule interacting with $U_{\alpha}$ 
\hl{and biased using the $\overline{Q}$ from $U_{\text{G\=o}}$} using 32 replicas, 
with a time compression of $\lambda_s=1$ (red), $\lambda_s=10$ (dark orange), $\lambda_s=100$ (light orange), and $\lambda_s=1000$ (yellow). Simulations are performed at $T=106$K starting from a conformation denatured at $400$K.}
\label{fig_q_protG_q}
\end{figure}

Given the ability of pMaxCal replica--simulation to modulate the dynamics of a simple system, we challenged the algorithm with a larger system. We defined two models for the full protein G B1 domain. The first is described by the standard G\=o potential $U_{\text{G\=o}}$ and the second in which the G\=o potential is modified strengthening the intra-helix interactions by a factor of 2 (we shall label the latter as $U_\alpha$). The equilibrium properties of the two models are similar (cf. Fig. S16 in the Supplementary Materials), but their folding dynamics, starting from a disordered conformation, is different (cf. the shapes of $\overline{Q}$ displayed as dark--grey and light--grey curves in Fig. \ref{fig_q_protG_q}). A simulation, carried out over 32 replicas, biasing the molecule interacting with the potential $U_\alpha$ to follow the dynamics of the mean fraction of native contacts $\overline{Q}$ of the molecule interacting with $U_{\text{G\=o}}$ is almost indistinguishable from the dynamics of its reference simulation when comparing the biasing variable (cf. the red curve in Fig. \ref{fig_q_protG_q} and Fig. S17 in the Supplementary Materials). Importantly, the \hl{time evolution} of other conformational variables, like the total RMSD, the gyration radius, the RMSD restricted to the two $\beta$--hairpins and to the whole $\beta$--sheet are very similar to those of the reference system (see Fig. \ref{fig_ort_protG_q} and Fig. S18 in the Supplementary Materials). 

As noted in the the Methods section, the current approach allows to modify the time--resolved behavior of a force--field making use of some external time--resolved information, this means nonetheless that one should be able to run simulations on the same time scale of the time--resolved information of interest. What happen if one rescales the time--scale of the time resolved information by a factor $\lambda_s$? This could in principle allow running short simulations and yet reproducing the long--time behavior of the system. 
\hl{This would means that we might not only employ the MaxCal to improve the quality of a force--field but also to boost, on average, the sampling of reactive trajectories.} 

To test the effect of the rescaling at least in ideal cases, we repeated the above simulations rescaling the time scale of the target reference--data by factors $\lambda_s=10$, $\lambda_s=100$ and $\lambda_s=1000$. In Figs. \ref{fig_q_protG_q} and \ref{fig_ort_protG_q} we compared the dynamics of the biasing coordinate and of some other coordinates, respectively (cf. also Figs. S17 and S18 in the Supplementary Materials), with that of the reference system interacting with $U_{\text{G\={o}}}$, rescaling back the time axis to the original time scale to allow a clear comparison. A rescaling factor $\lambda_s=10$ gives results which are essentially identical to the case without rescaling. With a rescaling factor $\lambda_s=100$, the qualitative agreement is still good, but the two curves are no longer perfectly overlapping, while a factor $\lambda_s=1000$ gives a dynamics which is completely different from both the unbiased and the reference--molecule ones (cf. also Fig. S18 in the Supplementary Materials). 

To study how the bias affects the different time scales of the dynamics of the model protein, we performed a time--lagged independent component analysis (TICA)\cite{Molgedey:1994ew,PerezHernandez:2013ih,Schwantes:2013bp} on the unbiased and on the biased simulations. This analysis combines information coming from covariance and time--lagged covariance matrix of the C$_\alpha$ positions, obtaining a qualitative estimate of the relaxation times of slow variables given a linear combination of trajectory observables (cf. Fig. S20 in the Supplementary Materials). The two original potentials $U_{\text{G\=o}}$ and $U_{\alpha}$ show significantly different relaxation times, and the caliber--biased simulation with $\lambda_{s}=1$ displays a good agreement with the reference potential relaxation times, demonstrating once again that replica--averaged time--resolved simulations could be used to include time--resolved data in MD. As expected, with the increase of $\lambda_{s}$ the system shows a speed up in all the slow variables. The worse behavior of the simulations with $\lambda_{s}=100$ and $1000$ can be explained considering the system diffusion time, which is in the order of 1 ps: With a too strong time rescaling, the resulting `slow' relaxation time is in the order of the ps, and thus the system cannot follow the bias (cf. Fig. S20 in the Supplementary Materials). 

\subsection{Biasing the dynamics using lower resolution observables}

All the former simulations have been biased to follow observables closely related to the reaction coordinate of the process (i.e. in this case protein folding). To test our approach in the case of more realistic observables, we used the same two models described in Sect. \ref{sect:protG} and used the ideal SAXS intensities as our source of synthetic information. We calculated the SAXS intensities from the reference system interacting with $U_{\text{G\=o}}$ and used the dynamics of the SAXS intensities at 15 equispaced values of the scattering vector as reference data to bias the model interacting with $U_\alpha$. 

\begin{figure}[htp]
\centering
\includegraphics[width=0.95\textwidth]{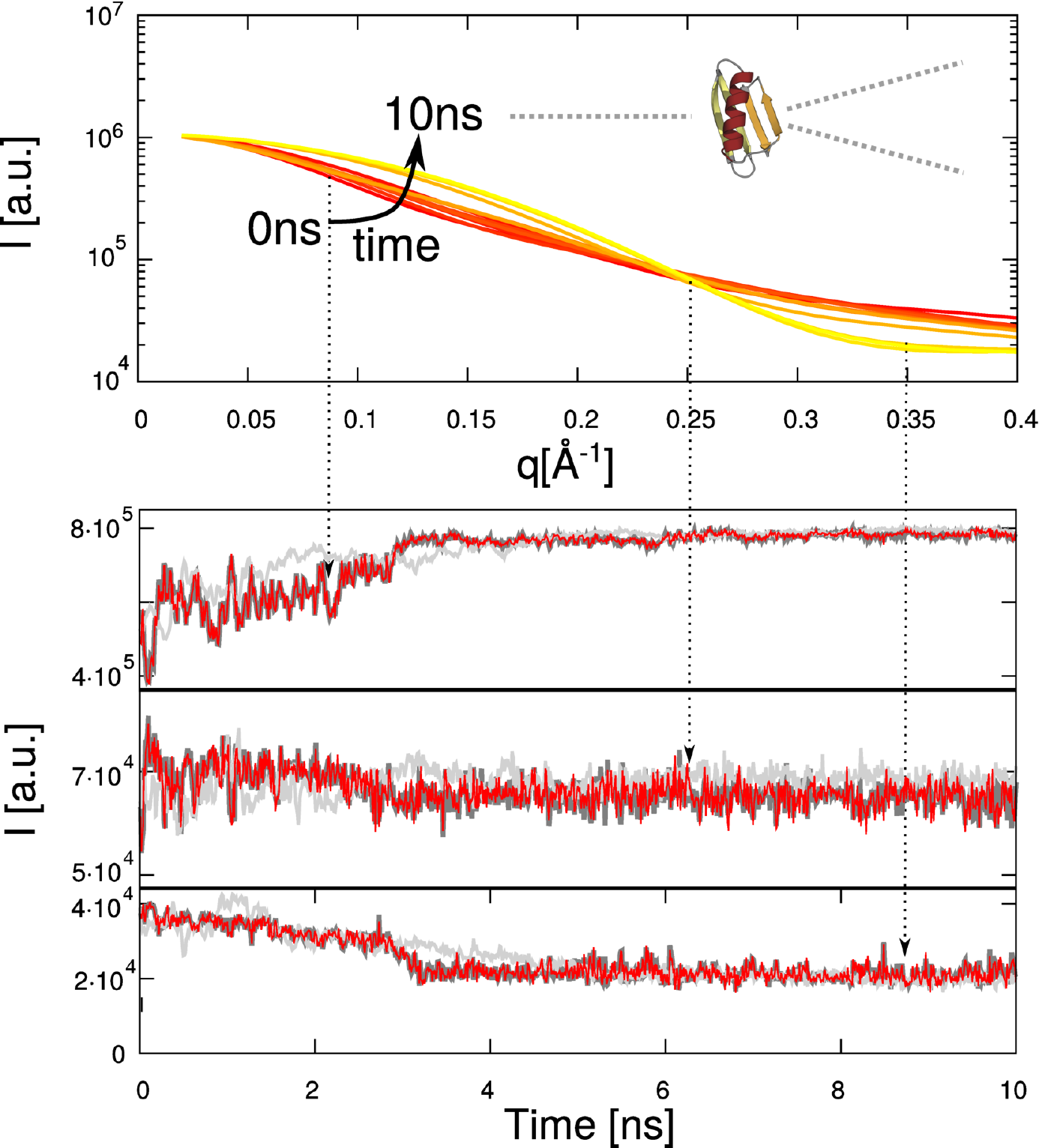}
\caption{In the upper panel, the time evolution of the SAXS spectrum simulated for the model of protein G interacting with  $U_{\text{G\=o}}$. In the lower panel, the evolution of the SAXS intensities at $q=0.08$\AA$^{-1}$, at $q=0.25$\AA$^{-1}$ and $q=0.35$\AA$^{-1}$. The light grey curve is the unbiased dynamics ($U_{\alpha}$), 
\hl{the dark-grey curve is the reference dynamics ($U_{\text{G\=o}}$) and the red curve is the time evolution for $U_{\alpha}$ biased using the SAXS intensities from $U_{\text{G\=o}}$.}}
\label{fig_saxs}
\end{figure}

The dynamics of the SAXS intensities obtained from the reference simulations is displayed in the upper panel of Fig. \ref{fig_saxs}, while in the lower panel it is shown the dynamics of the SAXS intensities at the values of $q$ ($0.08$\AA$^{-1}$, $0.25$\AA$^{-1}$ and $0.35$\AA$^{-1}$), chosen as an example. For these $q$ and for all the others (not shown here), the biased dynamics can follow perfectly well the dynamics of the reference system. In Fig. \ref{fig_ort_protG_saxs} it is shown the dynamics of  the radius of gyration, the RMSD of hairpins $\beta$1-2 and the native contact fraction, 
\hl{observables that are not used for biasing the simulation.} 
The biased simulations appear in good agreement with the reference dynamics (other conformational variables are shown in Fig. S19 in the Supplementary Materials). Finally, also the TICA-derived slow variables relaxation times are in good agreement with the ones of the unbiased reference potential (cf. Fig. S20 in the Supplementary Materials). Overall our simple--model calculations suggest that at least in principle it could be possible to integrate time--resolved data in MD simulations to modulate and possibly improve their agreement with some available knowledge.

\section{Discussion}

The quality of molecular mechanics force--fields is generally improving\cite{Beauchamp:2012eq,LindorffLarsen:2012gl,Tan:2018gv}, but these improvements, even if significant, are limited by the difficult of training force--fields on systems and or time--scales comparable to the one of interest. Hybrid, inferential, methods based on the introduction of equilibrium experimental information in MD simulations, either as an {\it a posteriori} reweighing or as a direct bias of the simulation\cite{Bonomi:2017dn}, can alleviate these limitations in a system dependent manner. Among these, replica--averaged simulations\cite{LindorffLarsen:2005bi}, based on the maximum entropy principle\cite{Cavalli:2013jf} and recently extended to include a Bayesian treatment of the errors\cite{Hummer:2015ba,Bonomi:2016ip}, have been particularly successful\cite{Boomsma:2014br,Bonomi:2017dn}. 

Inferential methods could also be used to integrate time--resolved informations. Here we showed that the principle of maximum caliber, previously used only to perform {\it a posteriori} reweighing\cite{Wan:2016fz,Zhou:2017bh,Dixit:2018hg,Dixit:2018iy}, can be implemented as a direct bias using a replica--averaged time--resolved MD scheme and that at least for simple--model systems can be used to modulate the behavior of time--resolved observables. 
\hl{Formally our current proof is valid for a uniform prior and a Brownian dynamics (cf. Section II), nonetheless the simulations suggests its general validity when a prior force--field is known and trajectories are obtained by MD.} 
Future works should also consider the effect of errors in the data that is currently missing (cf. \cite{Presse:2013dh}) and other forms of experimental informations like path-based information (cf. \cite{Touchette:2013gs}). 

Importantly, we have also tested the effect of rescaling the time--scale of the employed time--resolved data. Real--time experiments (H/D exchange\cite{Rand:2014bz}, real-time NMR\cite{vanNuland:1998iz} as well as time-resolved SAXS/WAXS\cite{Cammarata:2008cw,Pollack:2011hz}) are often employed to study processes on time scales that are longer than those usually accessible by MD (i.e. on the order of hundreds of microseconds to milliseconds and longer). In this case the choice of the biasing variable plays an important role to ensure the realism of the resulting trajectories. Our simple--models suggest that it is in principle possible to rescale the time--units of the data employed as long as this is longer than the diffusion time. Nonetheless more work is needed in this direction to assess specific observables. We anticipate that for observables correlated with the slowly--varying reaction coordinate of a system (like for the sand--box simulations described in Sects. \ref{sect:hairpin} and \ref{sect:protG}), the macroscopic dynamic will be correct even in case of strong rescaling while for observables weakly correlated with the reaction coordinate of the process the macroscopic dynamics of the system will mostly rely on the force field.

\begin{acknowledgements}
We thank Giovanni Bussi, Stefano Gianni and Toni Giorgino for useful discussion.
\end{acknowledgements}

\bibliography{caliber}

\newpage

\newpage

\newpage
\begin{figure}[htp]
\centering
\includegraphics[width=0.9\textwidth]{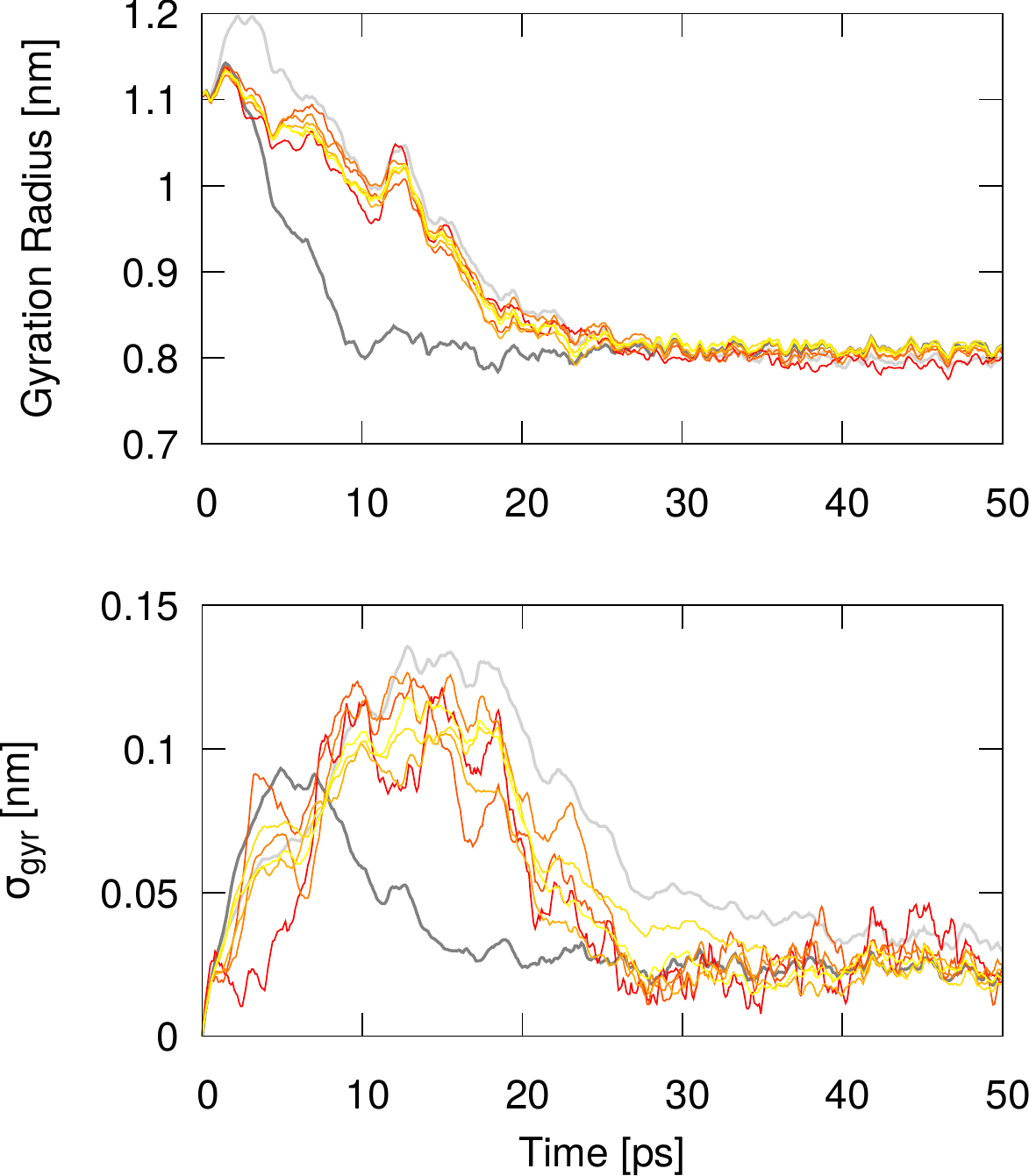}
\caption{The time evolution of the gyration radius (top) and its fluctuations (bottom) of the hairpin. The dark--grey line indicate the dynamics generated with $U_{\text{head}}$, the light--grey line is the reference dynamics generated with $U_{\text{tail}}$ and the colored lines are the simulations performed with $U_{\text{head}}$ and 
\hl{biased using the $\overline{Q}$ from $U_{\text{tail}}$ (cf. Fig. 2)} 
using from 4 (red) to 128 replicas (yellow).}
\label{chi_q_tc_other}
\end{figure}

\newpage

\newpage
\begin{figure}[htp]
\centering
\includegraphics[width=0.95\textwidth]{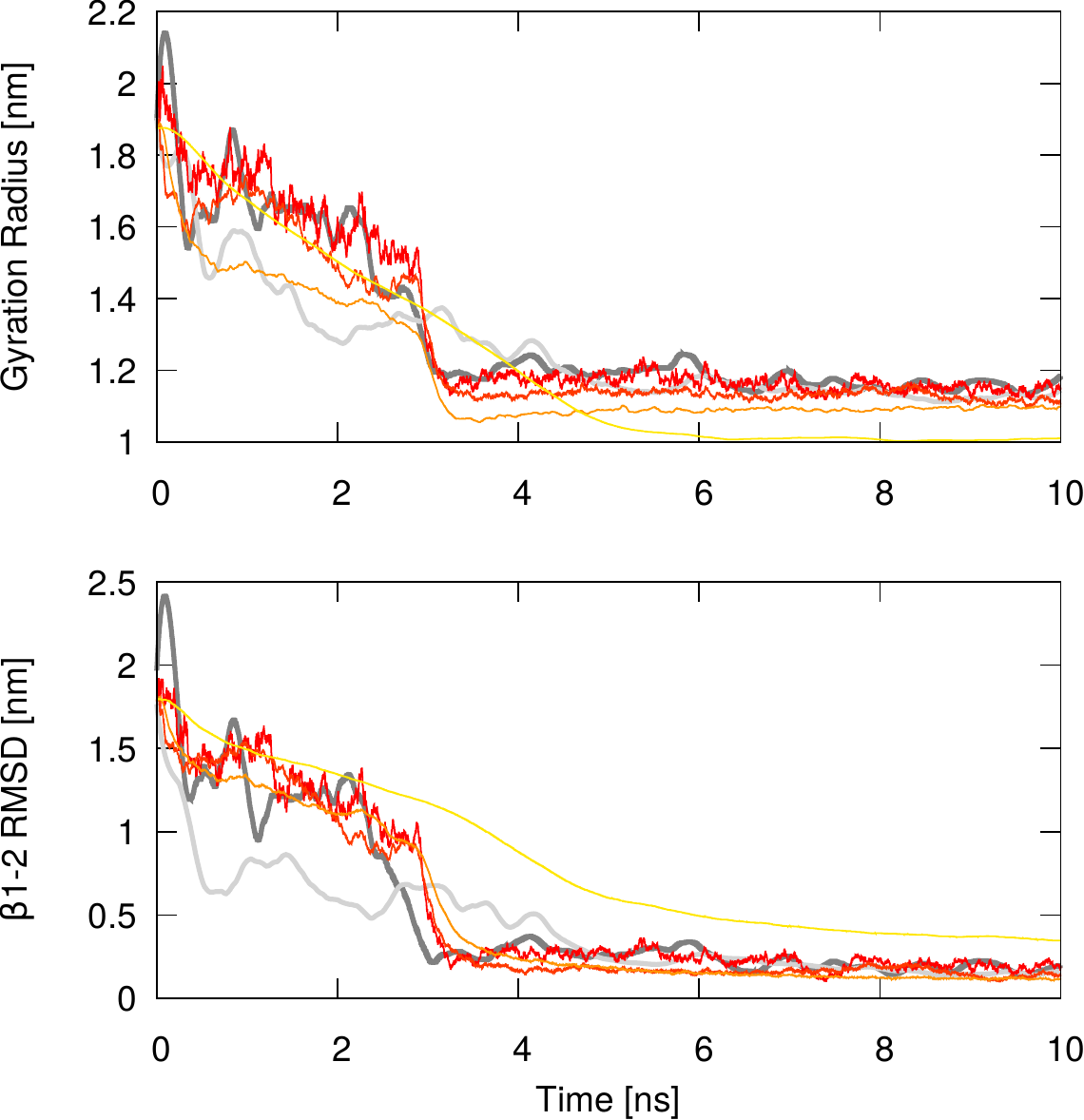}
\caption{The time evolution of the average gyration radius (top) and the RMSD of the interface between $\beta$-hairpins 1-2 (bottom) for the same simulations displayed in Fig. \protect\ref{fig_q_protG_q}.  The unbiased simulations of protein G interacting with $U_{\text{G\=o}}$ (dark grey); the unbiased simulations interacting with $U_{\alpha}$ (light grey); and the three biased simulations of the molecule interacting with $U_{\alpha}$ 
\hl{and biased using the $\overline{Q}$ from $U_{\text{G\=o}}$}
using 32 replicas, with a time compression of $\lambda_s=1$ (red), $\lambda_s=10$ (dark orange), $\lambda_s=100$ (light orange), and $\lambda_s=1000$ (yellow) (cf. Fig. 4).}
\label{fig_ort_protG_q}
\end{figure}

\newpage

\newpage
\begin{figure}[htp]
\centering
\includegraphics[width=0.95\textwidth]{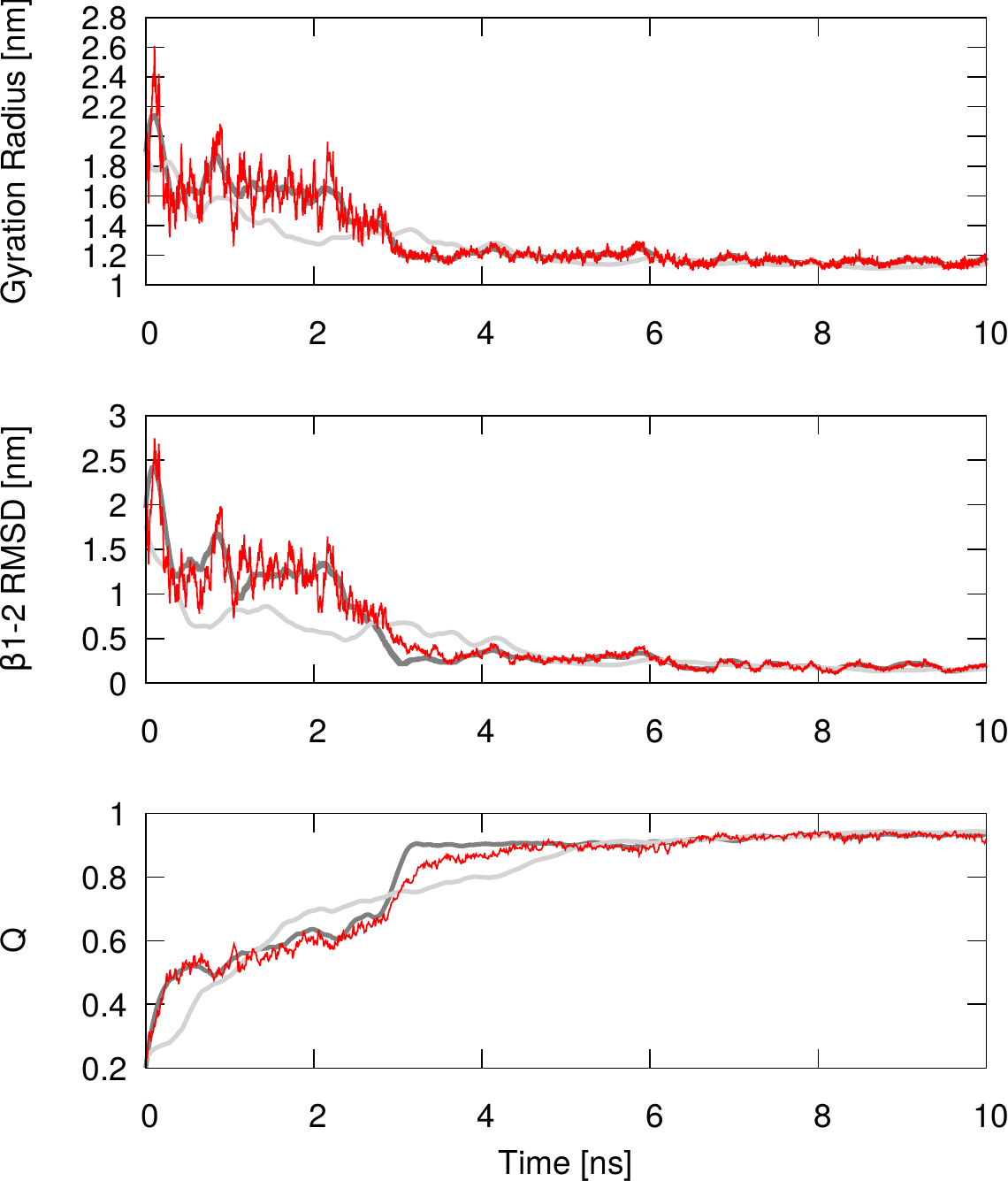}
\caption{The time evolution of gyration radius (top), RMSD of the interface between $\beta$-hairpins 1-2 (middle), and fraction of native contacts ($Q$) of protein G obtained biasing by means of the ideal SAXS intensities. 
\hl{the dark-grey curve is the reference dynamics ($U_{\text{G\=o}}$) and the red curve is the time evolution for $U_{\alpha}$ biased using the SAXS intensities from $U_{\text{G\=o}}$ (cf. Fig. 6)}}
\label{fig_ort_protG_saxs}
\end{figure}

\newpage
\appendix {\bf Supplementary Figures}
\renewcommand{\thefigure}{S\arabic{figure}}
\setcounter{figure}{0}
\begin{figure}[htp]
\includegraphics[width=0.8\textwidth]{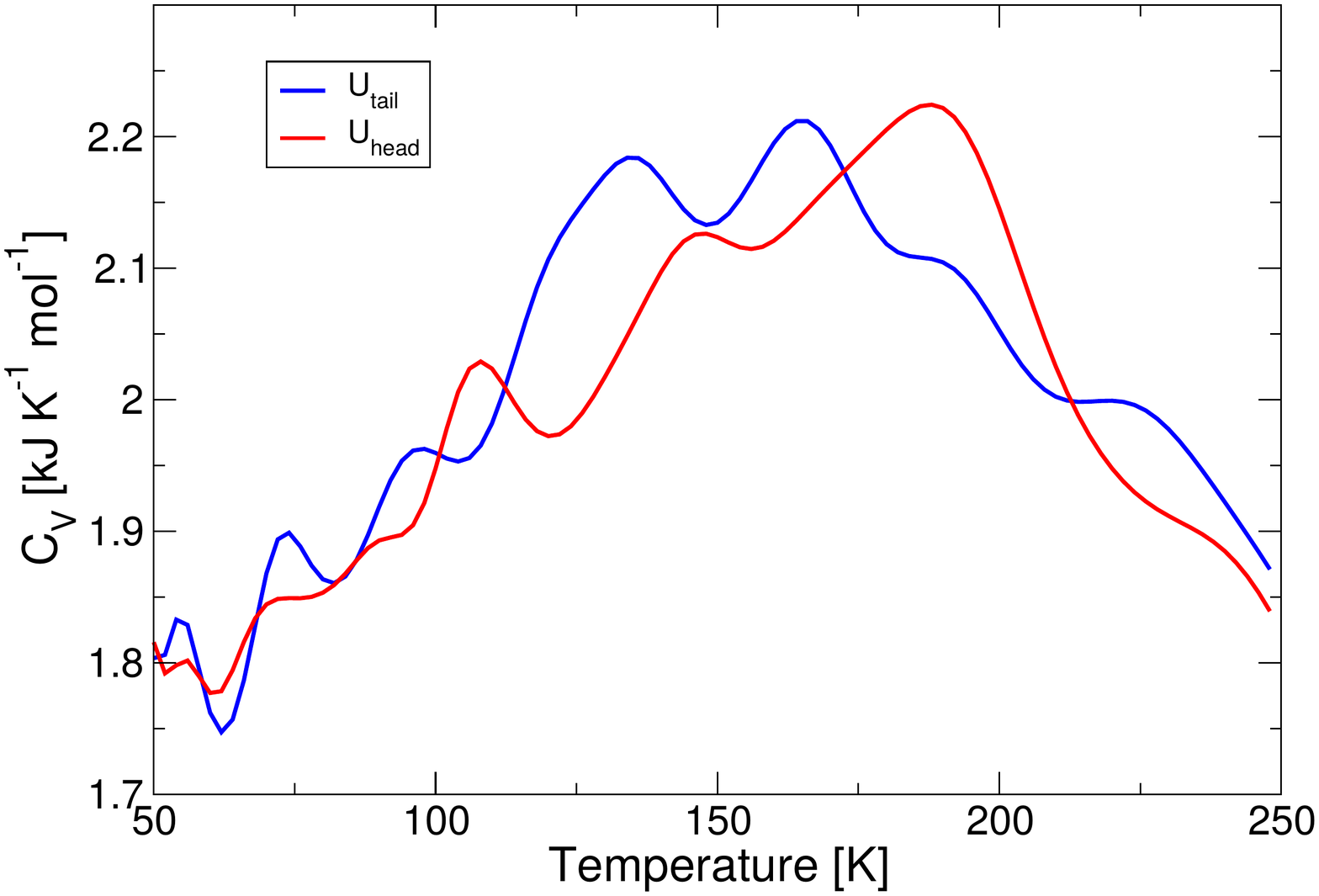}
\caption{Heat capacity as a function of the  temperature for the hairpin of protein G interacting with  $U_{\text{tail}}$ (blue) and $U_{\text{head}}$ (red). The curves are obtained from replica--exchange simulations using 25 replicas with temperatures from 50 to 250K and analyzed with a weighted--histogram algorithm.}
\label{hairpin_cv}
\end{figure}

\begin{figure}[htp]
\includegraphics[width=0.9\textwidth]{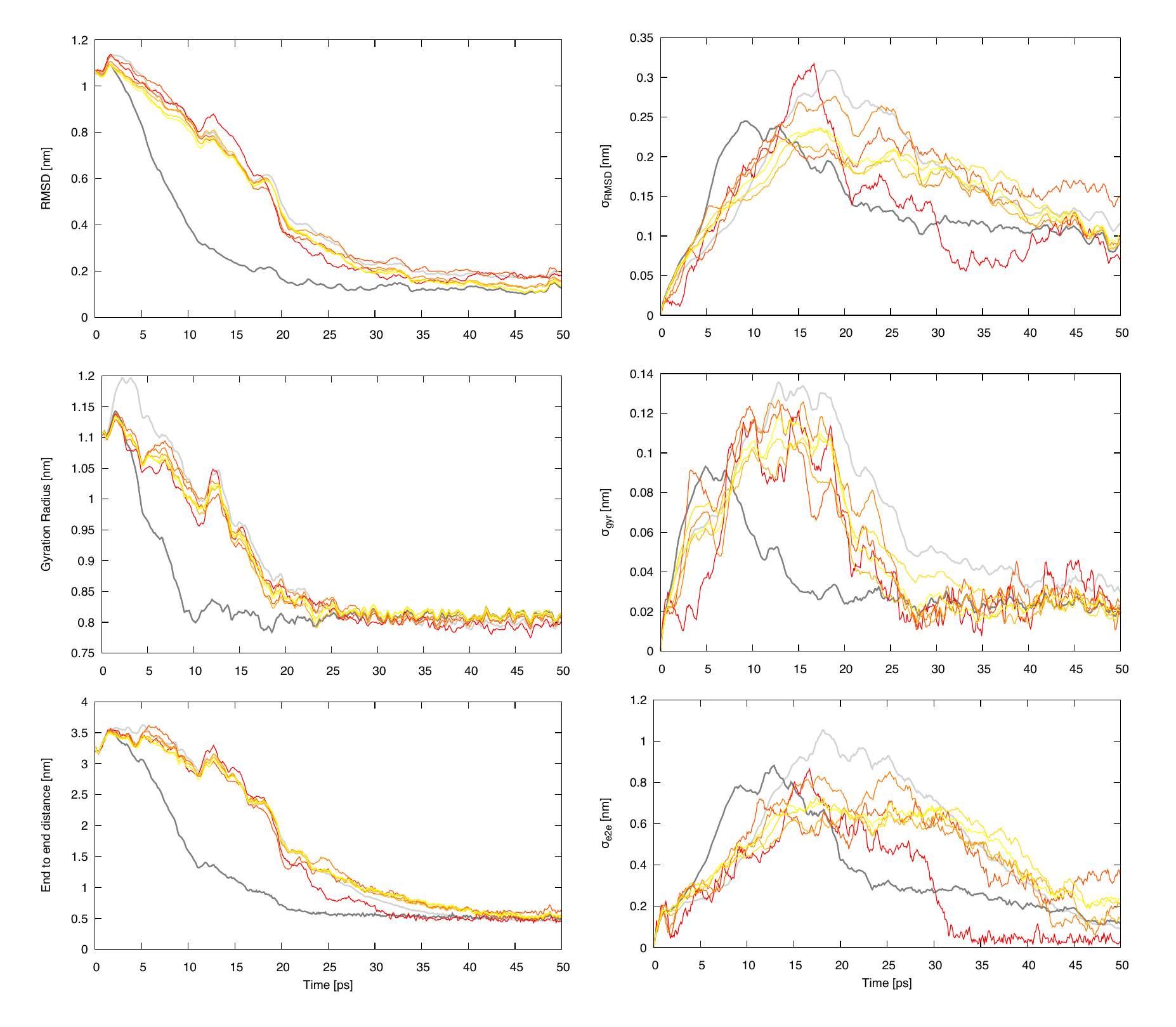}
\caption{To the left, the time-evolution of the RMSD (top), the gyration radius (middle) and the end--to--end distance (bottom) of the hairpin model system (cf. Fig. 2). The dark--grey line indicates the time evolution generated with $U_{\text{head}}$, the light--grey line is the reference generated with $U_{\text{tail}}$ and the colored lines are the simulations performed with $U_{\text{head}}$ and biased by the $\overline{Q}$ from $U_{\text{tail}}$ using from  4 (red) to 128 replicas (yellow). To the right, the standard deviations over the replicas of the same quantities.}
\label{chi_q_tc_other}
\end{figure}

\begin{figure}[htp]
\includegraphics[width=0.7\textwidth]{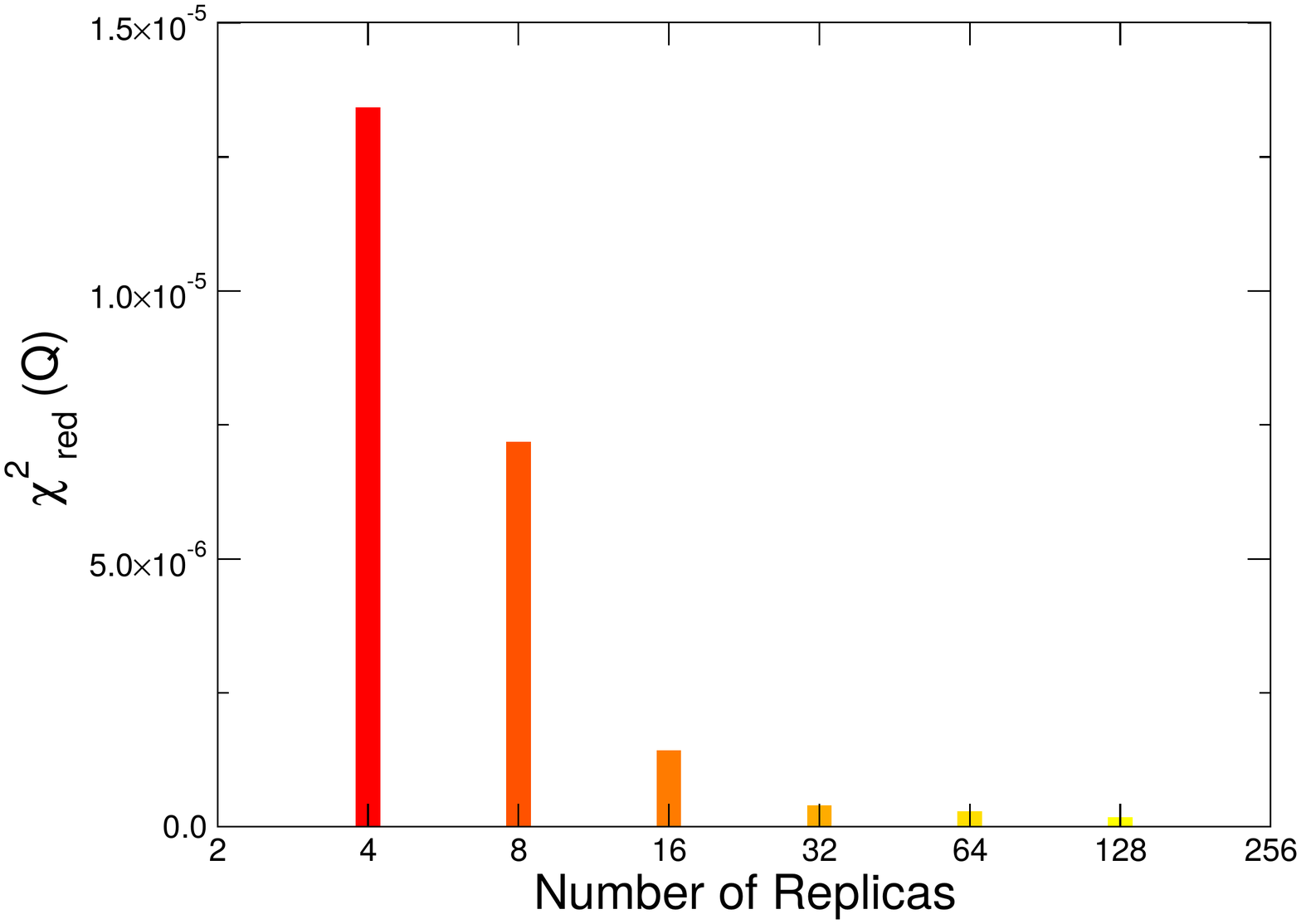}
\caption{The $\chi^{2}_{\text{red}}$, defined as $\chi_{\text{red}}^{2} = \frac{1}{N} \sum_{t} \frac{({\overline Q}^{\text{bias}(t)} - {\overline Q}^{\text{ref}}(t))^{2}}{\overline{Q}^{\text{ref}(t)}}$, between the points of the function ${\overline Q}$ of the system interacting with $U_{\text{head}}$ and biased in simulations with a variable number of replicas and that of the system interacting with $U_{\text{tail}}$, regarded as the reference system.}
\label{chi_q_tc}
\end{figure}

\begin{figure}[htp]
\includegraphics[width=0.9\textwidth]{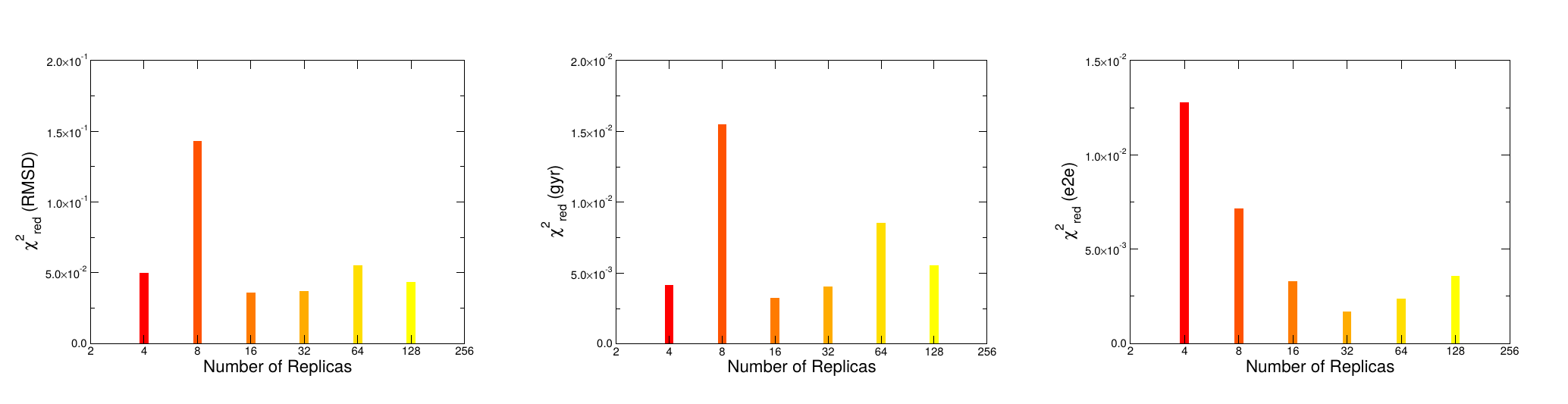}
\caption{The $\chi^{2}_{\text{red}}$ (defined as in Fig. \protect\ref{chi_q_tc})  for the curves displayed in Fig. S2.}
\label{chi_ort_tc_fluct}
\end{figure}

\begin{figure}[htp]
\includegraphics[width=0.7\textwidth]{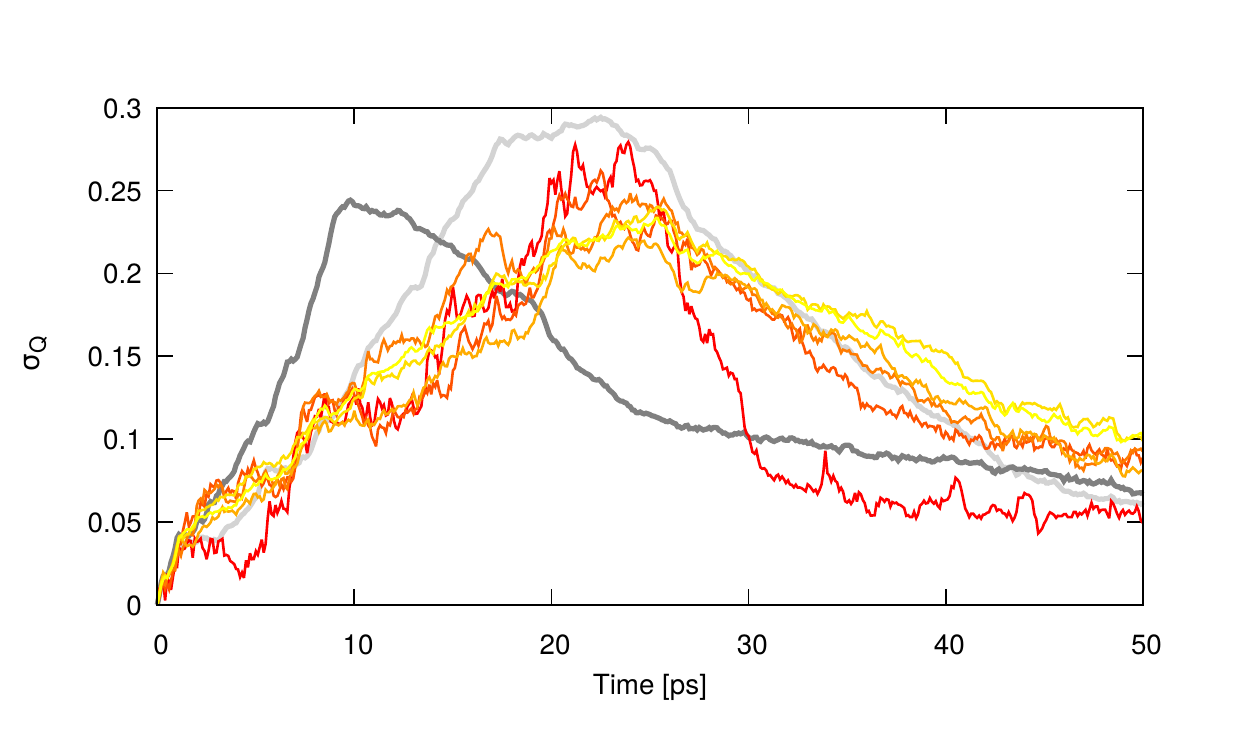}
\caption{The standard deviation of $Q$ over the replicas as a function of time for the unbiased system interacting with the potential $U_{\text{head}}$ is displayed in dark grey. The light grey curve is obtained from the system interacting with the potential $U_{\text{tail}}$, while colored solid lines are those obtained applying the caliber to simulations of the system interacting with $U_{\text{head}}$ through a biasing potential depending on $\overline{Q}$ with a number of replicas from 4 (red) to 128  (yellow).}
\label{img:fig_q_tc_fluct}
\end{figure}

\begin{figure}[htp]
\includegraphics[width=0.7\textwidth]{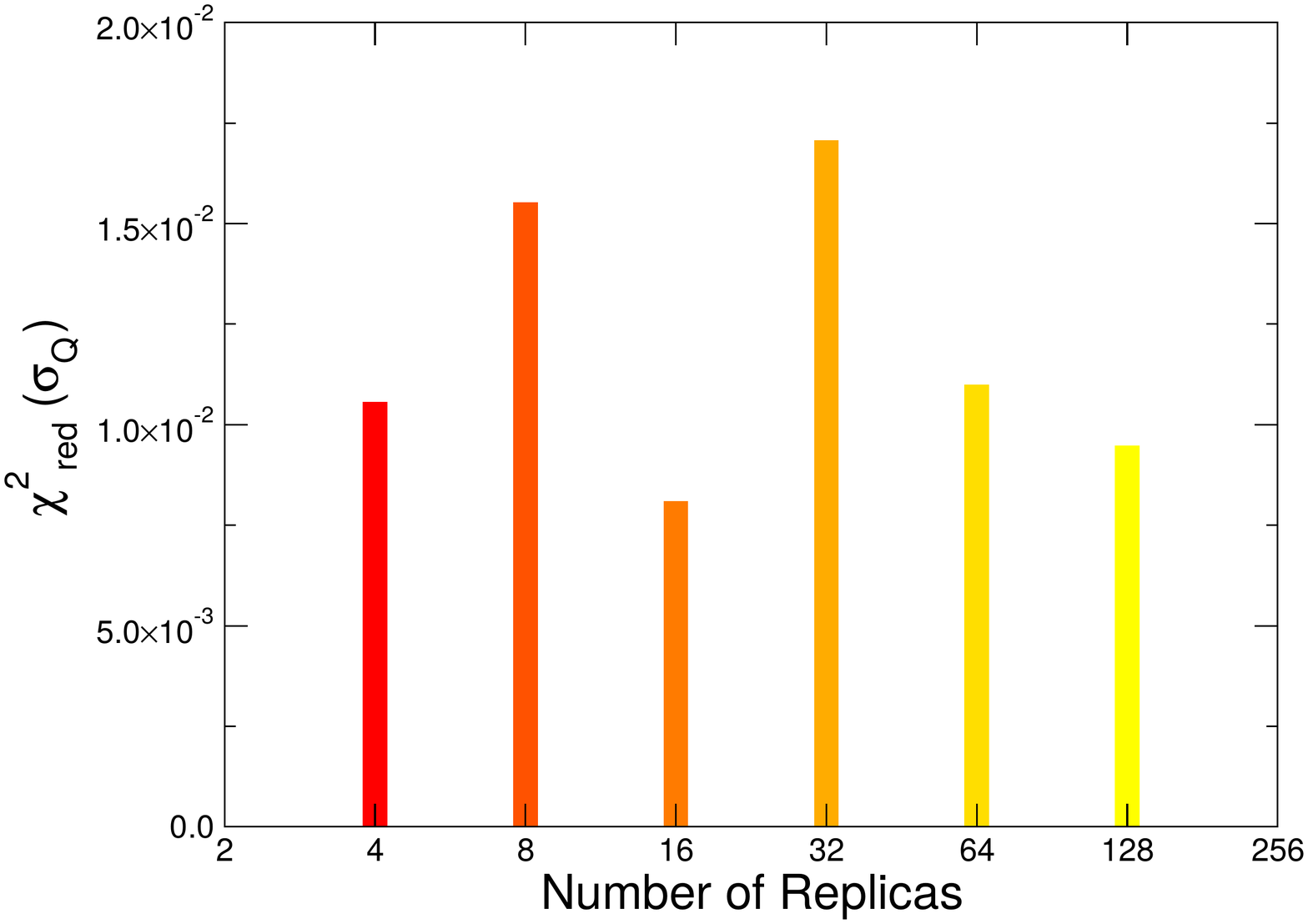}
\caption{The $\chi^{2}_{\text{red}}$ (defined as in Fig. \protect\ref{chi_q_tc})  for the biased curves (with a number of replicas from 4 (red) to 128  (yellow)) displayed in Fig. \protect\ref{img:fig_q_tc_fluct}.}
\label{img:chi_q_tc_fluct}
\end{figure}

\begin{figure}[htp]
\includegraphics[width=0.7\textwidth]{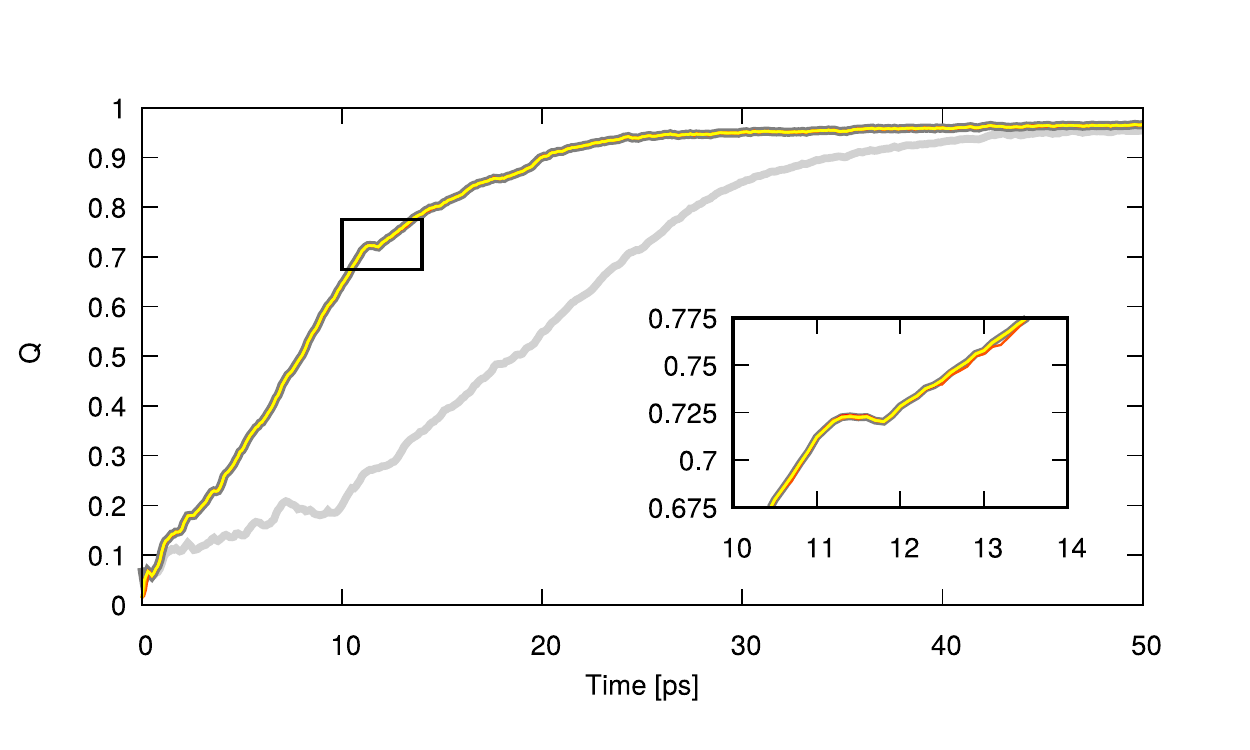}
\caption{The average fraction of native contacts obtained biasing the system interacting with $U_{\text{tail}}$ (light grey curve) to display the same time evolution as that interacting with  $U_{\text{head}}$ (dark grey curve). The colored curves indicate the biased trajectories (cf. the caption of Fig. 2).}
\label{img:fig_q_ct}
\end{figure}

\begin{figure}[htp]
\includegraphics[width=0.7\textwidth]{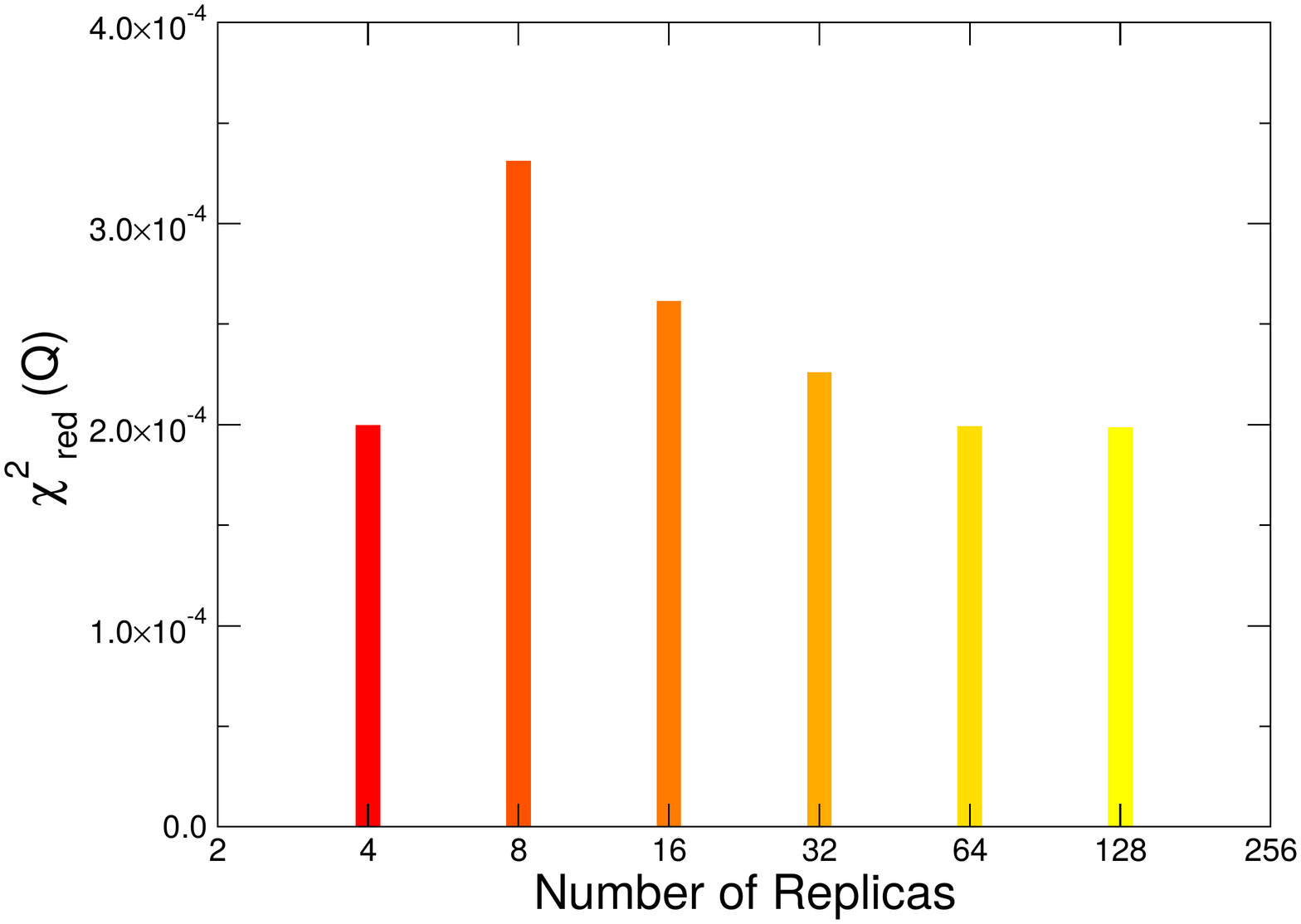}
\caption{The $\chi^2$ between the time evolution of $\bar{Q}$ obtained from biased and target trajectories (cf. Fig. \protect\ref{chi_q_tc}), obtained biasing the system interacting with $U_{\text{tail}}$ to display the same dynamics as that interacting with  $U_{\text{head}}$, thus quantifying the difference between the curves displayed in Fig. \protect\ref{img:fig_q_ct}.}
\label{img:chi_q_ct}
\end{figure}

\begin{figure}[htp]
\includegraphics[width=0.7\textwidth]{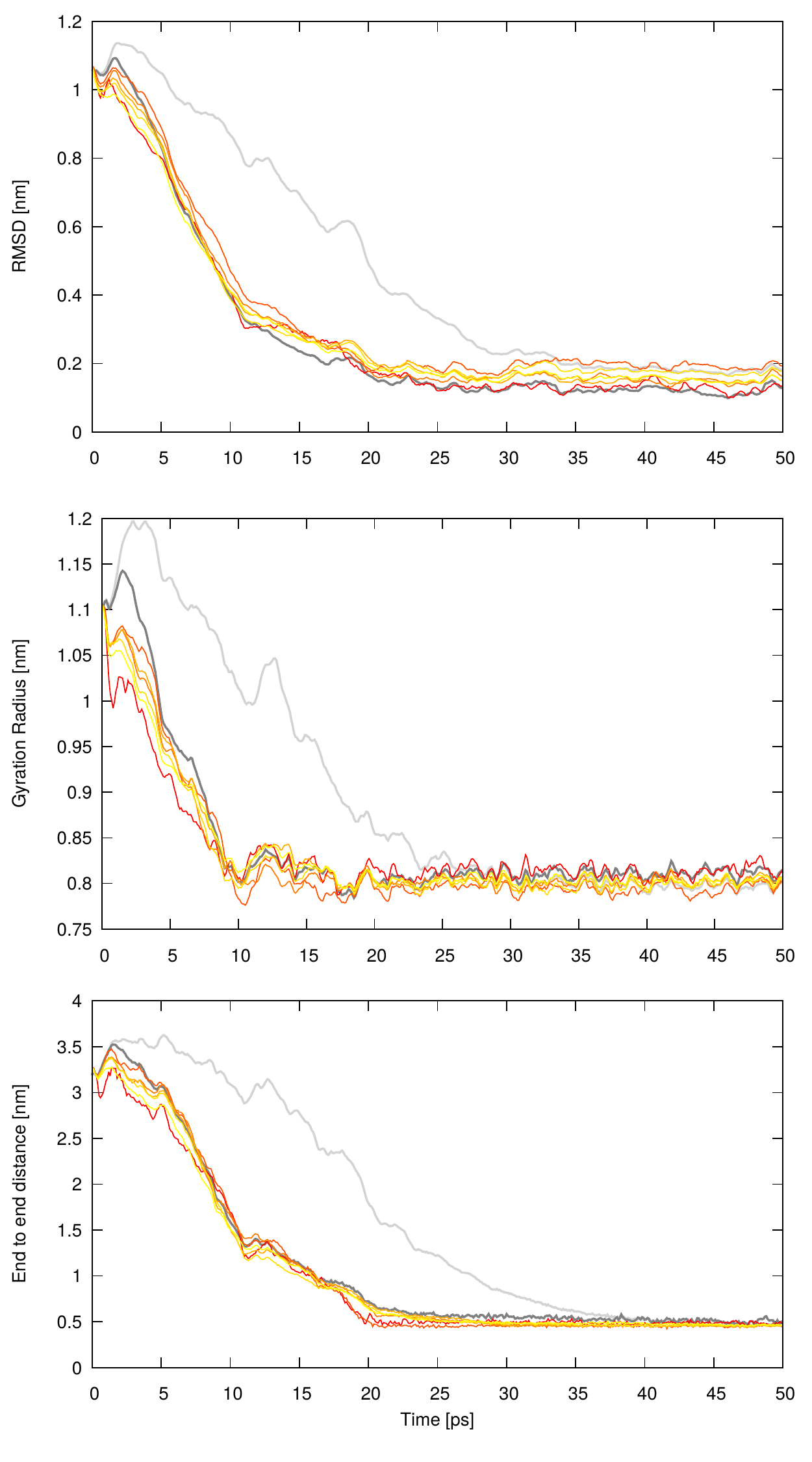}
\caption{Time evolution of variables different than $\bar{Q}$, obtained biasing the system interacting with $U_{\text{tail}}$ to display the same dynamics as that interacting with  $U_{\text{head}}$. The colored curves indicate the biased trajectories (cf. the caption of Fig. 2).}
\label{img:fig_ort_ct}
\end{figure}

\begin{figure}[htp]
\centering
\includegraphics[width=0.95\textwidth]{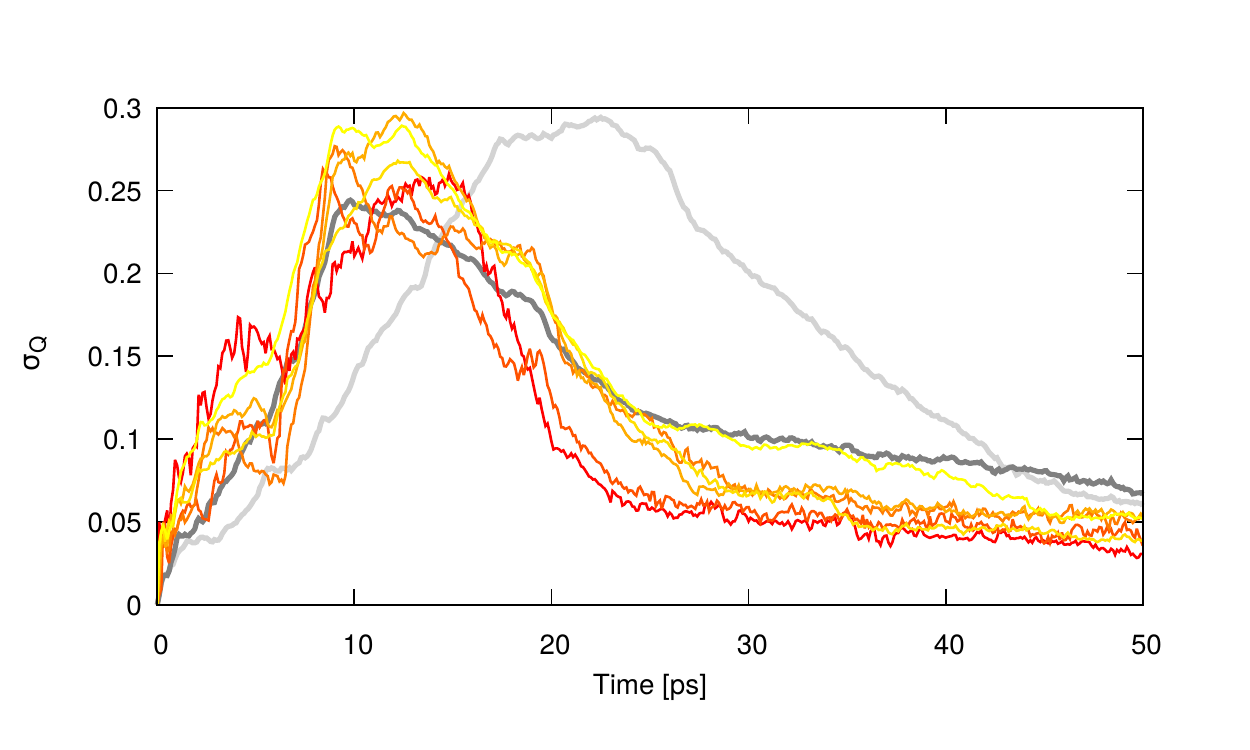}
\caption{Time evolution of the standard deviation of $Q$ (cf. Fig. \protect\ref{img:fig_q_tc_fluct}), obtained biasing the system interacting with $U_{\text{tail}}$ to display the same dynamics as that interacting with  $U_{\text{head}}$. The colored curves indicate the biased trajectories (cf. the caption of Fig. 2).}
\label{img:fig_q_ct_fluct}
\end{figure}

\begin{figure}[htp]
\centering
\includegraphics[width=0.7\textwidth]{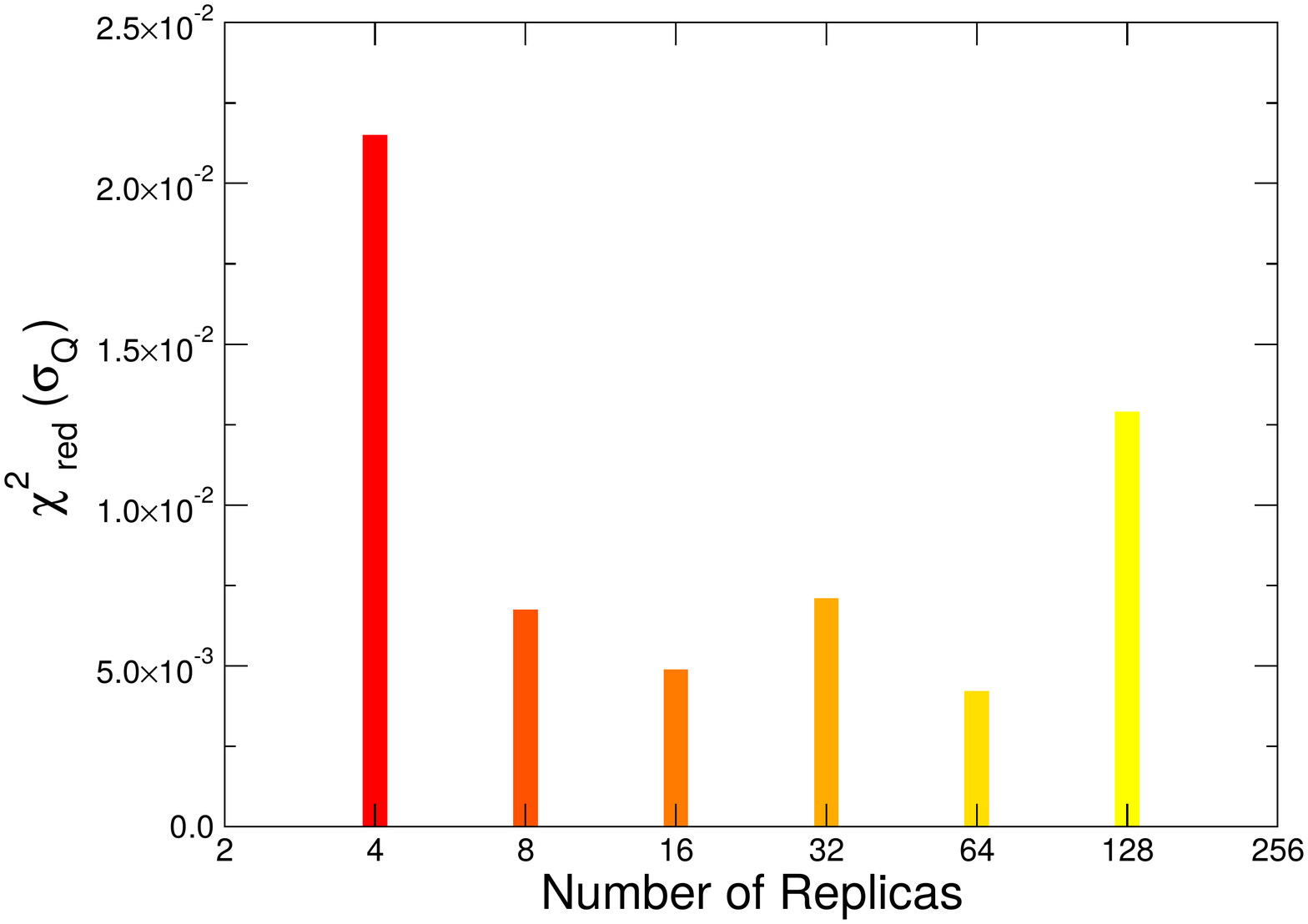}
\caption{The $\chi^{2}_{\text{red}}$ between the curves displayed in Fig. \protect\ref{img:fig_q_ct_fluct}}
\label{img:chi_q_ct_fluct}
\end{figure}

\begin{figure}[htp]
\centering
\includegraphics[width=0.55\textwidth]{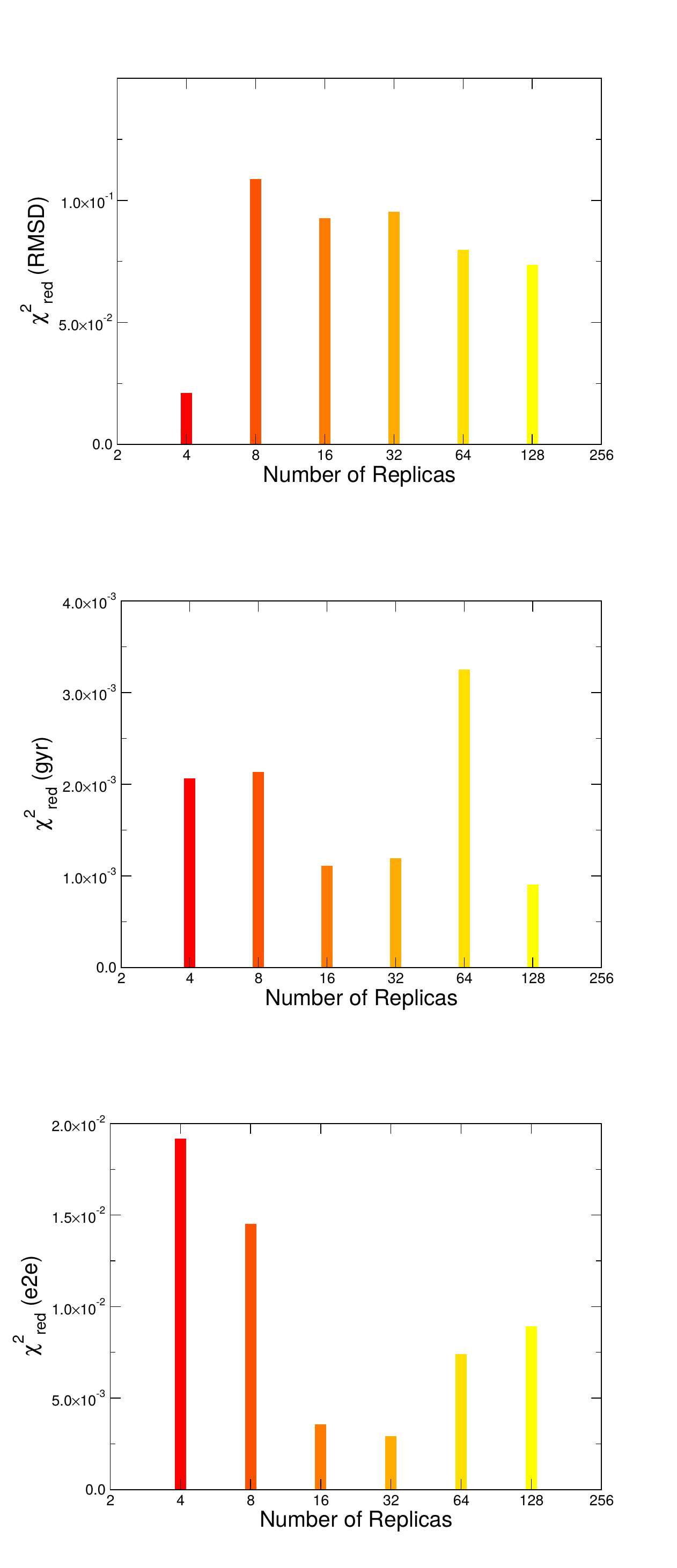}
\caption{The $\chi^{2}_{\text{red}}$ between the curves displayed in Fig. \protect\ref{img:fig_ort_ct}}
\label{img:chi_ort_ct}
\end{figure}

\begin{figure}[htp]
\centering
\includegraphics[width=0.7\textwidth]{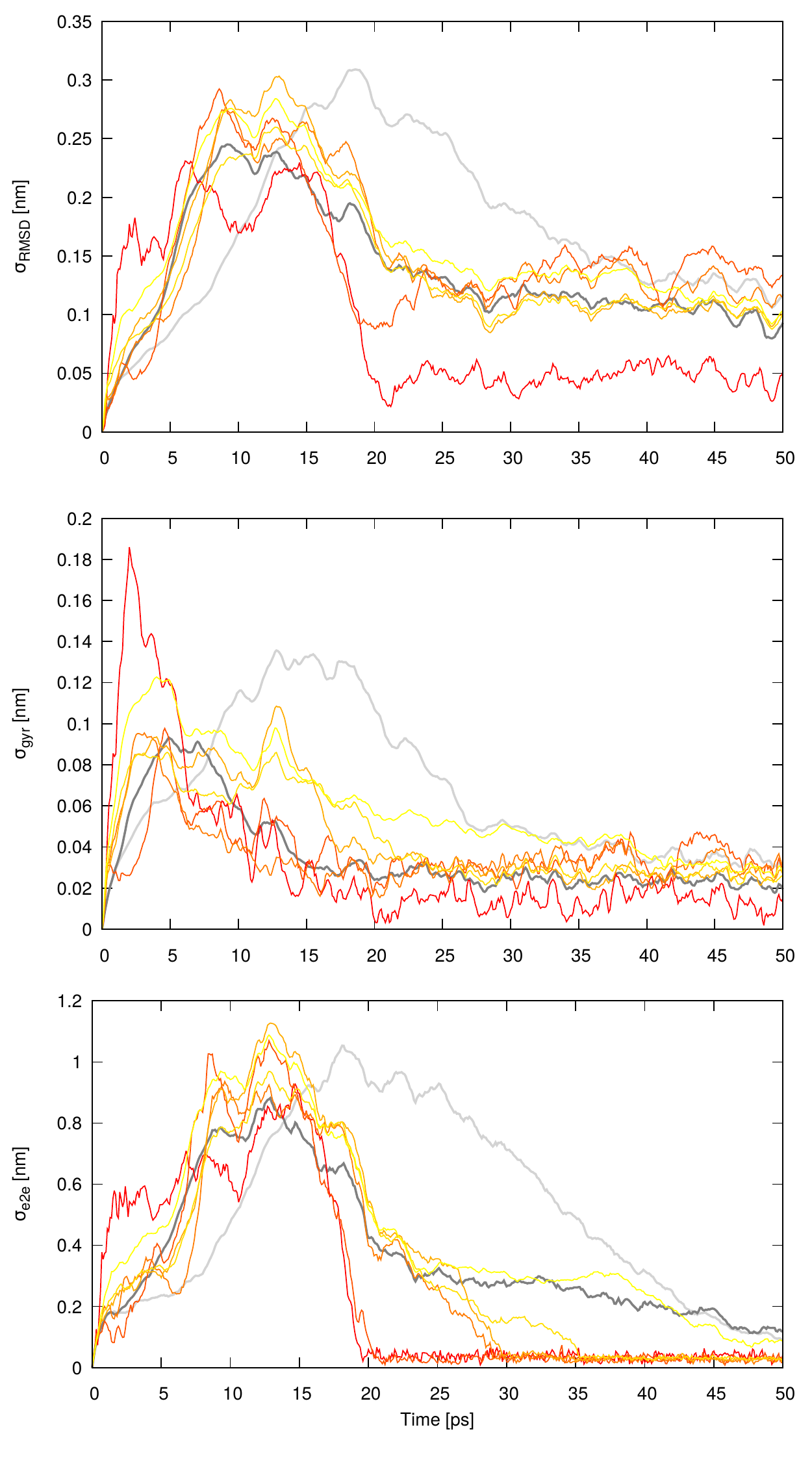}
\caption{The standard deviations associated with the averages displayed in Fig. \protect\ref{img:fig_ort_ct}}
\label{img:fig_ort_ct_fluct}
\end{figure}

\begin{figure}[htp]
\centering
\includegraphics[width=\textwidth]{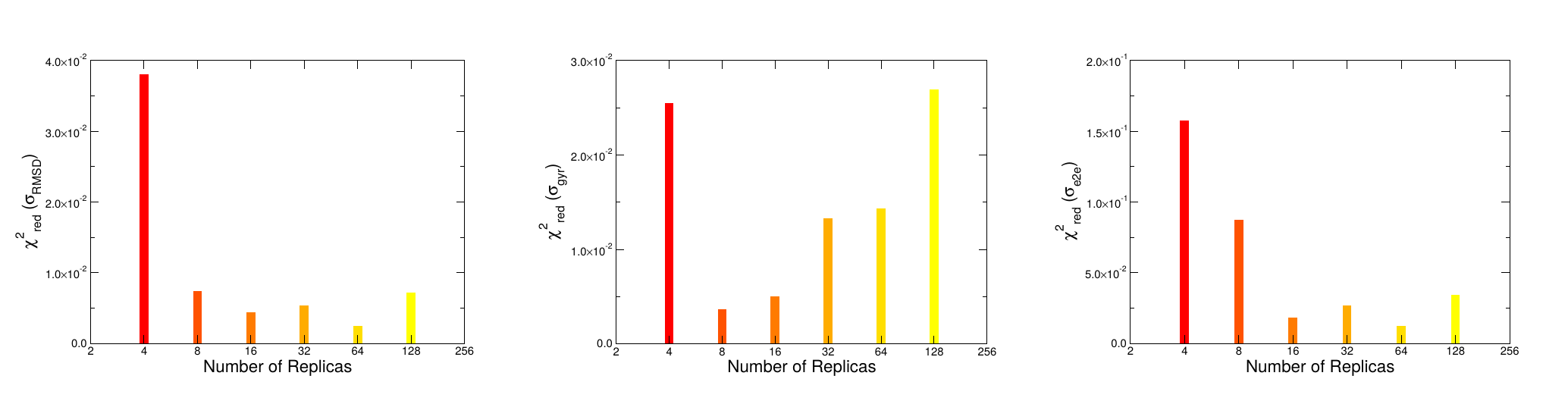}
\caption{The $\chi^{2}_{\text{red}}$ between the curves displayed in Fig. \protect\ref{img:fig_ort_ct_fluct}}
\label{img:chi_ort_ct_fluct}
\end{figure}

\begin{figure}[htp]
\centering
\includegraphics[width=0.95\textwidth]{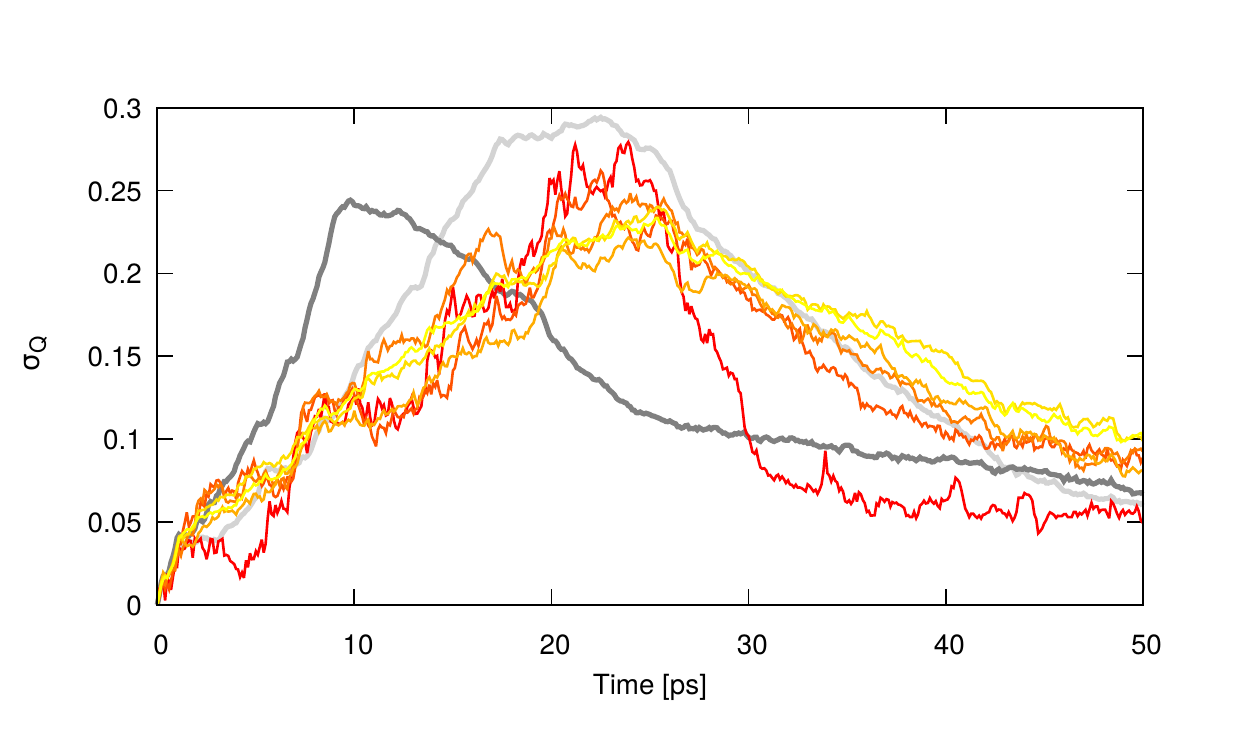}
\caption{Fluctuations of average fraction of native contacts ($Q$) as a function of time for unbiased $U_{\text{head}}$ (dark grey), unbiased $U_{\text{tail}}$ (light grey), and caliber restrained simulations from $U_{\text{head}}$ to $U_{\text{tail}}$, from 4 (red) to 128 replicas (yellow) in color scale.}
\label{fig_q_tc_fluct}
\end{figure}


\begin{figure}[htp]
\centering
\includegraphics[width=0.8\textwidth]{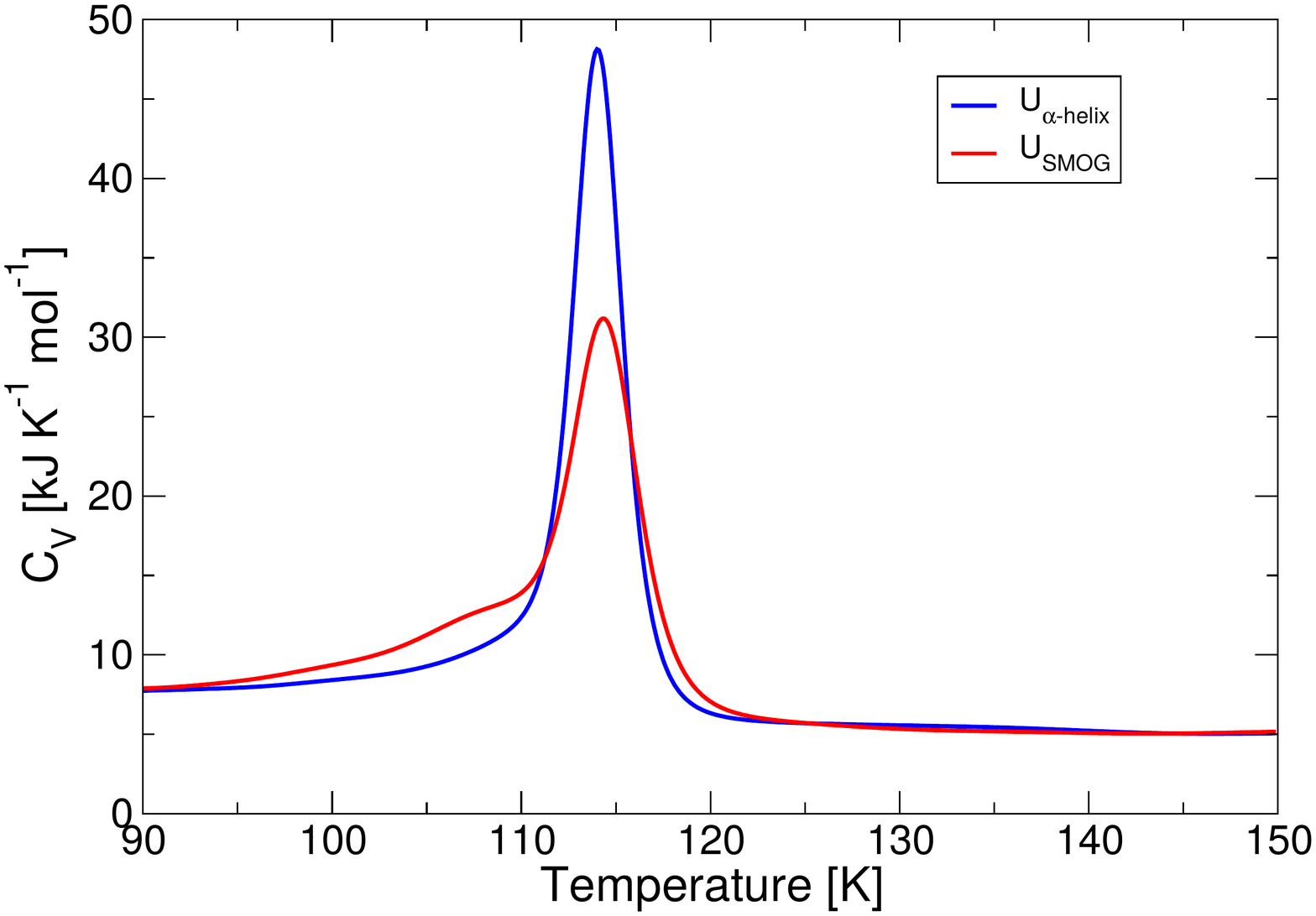}
\caption{Heat capacity in function of temperature for protein G under $U_{\alpha}$ (blue) and $U_{\text{G\=o}}$ (red) calculated from replica--exchange simulations through a multiple--histogram algorithm.}
\label{protG_cv}
\end{figure}

\begin{figure}[htp]
\centering
\includegraphics[width=0.9\textwidth]{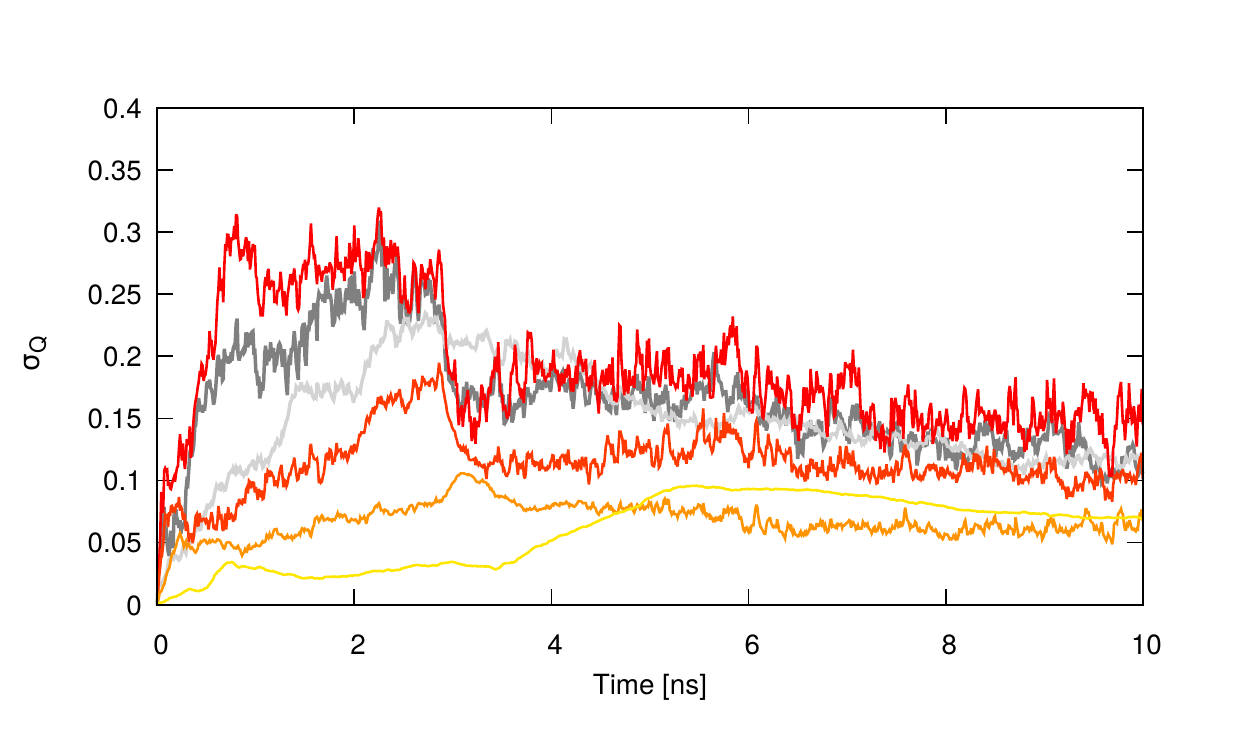}
\caption{The fluctuations over replicas of $Q$ in the simulation described in Fig. 4.}
\label{img:fig_q_protG_q_fluct}
\end{figure}

\begin{figure}[htp]
\centering
\includegraphics[width=0.95\textwidth]{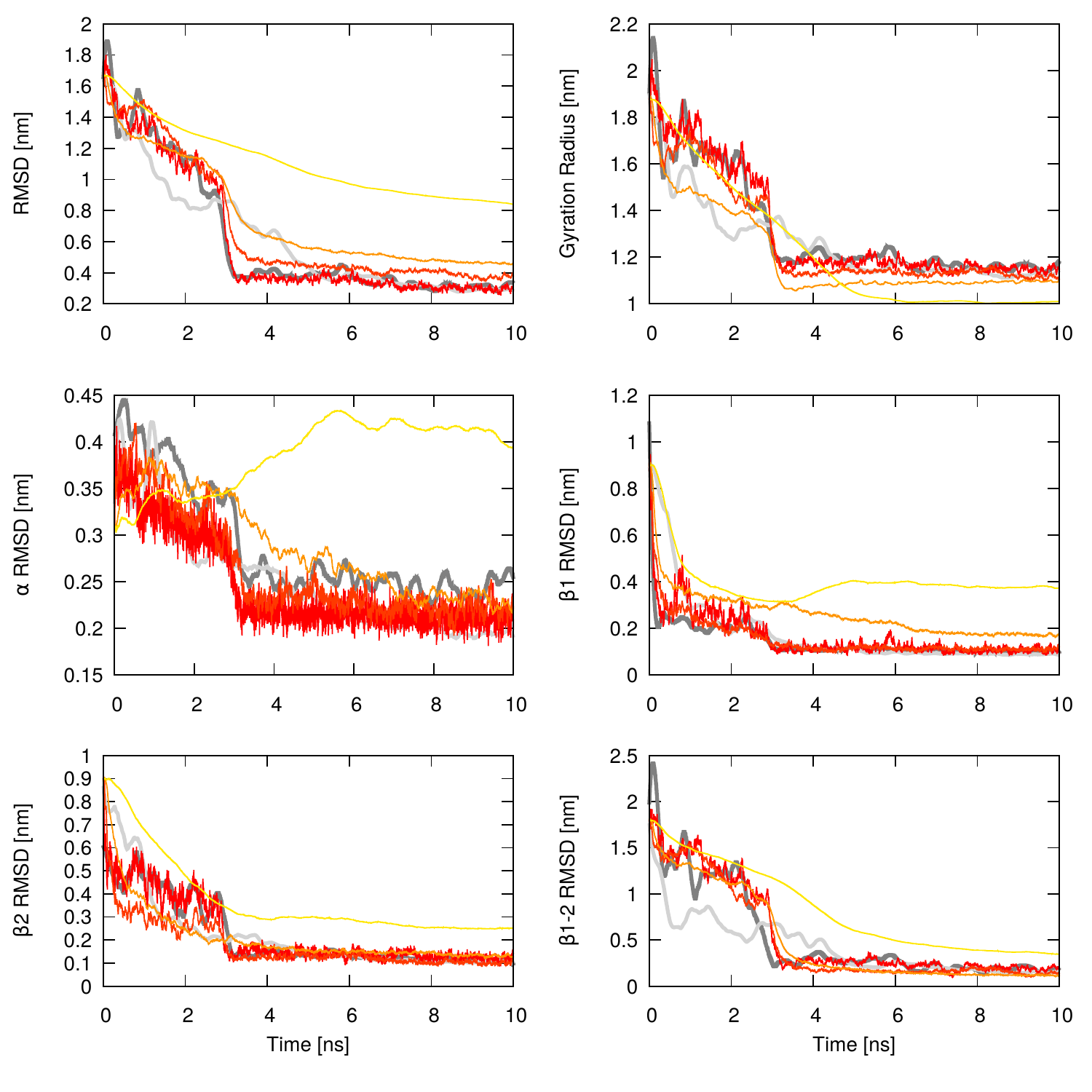}
\caption{The time evolution of the average C$_{\alpha}$-RMSD (top left), gyration radius (top right), $\alpha$-helix RMSD (center left), $\beta$-hairpin--1 RMSD (center right), $\beta$-hairpin--2 RMSD (bottom left), and the RMSD of the interface between $\beta$-hairpins 1-2 (bottom right) for the same simulations (and with the same color code) as those displayed in Fig. 4.}
\label{fig_ort_protG_q}
\end{figure}

\begin{figure}[htp]
\centering
\includegraphics[width=0.95\textwidth]{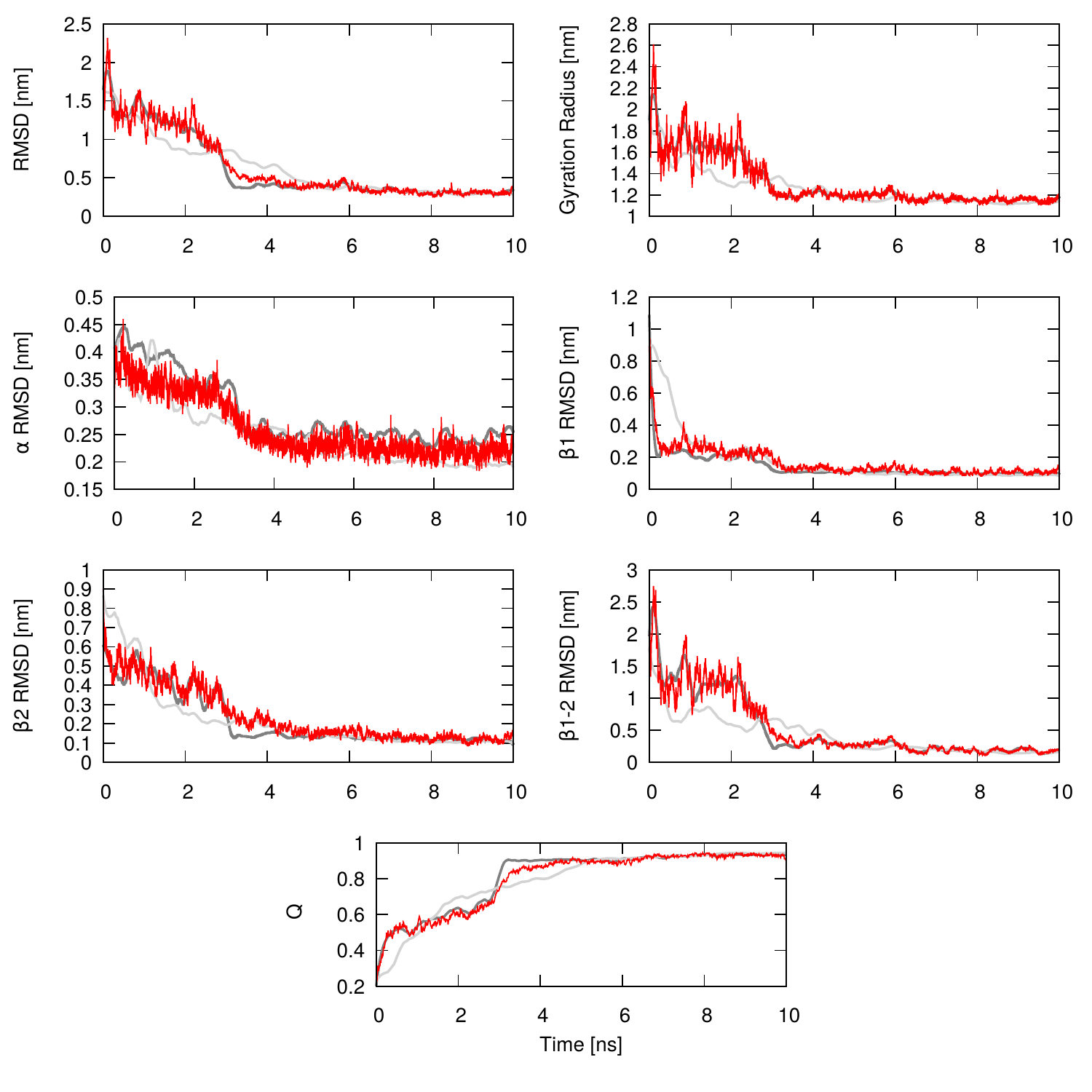}
\caption{The time evolution of some conformational coordinates of protein G obtained biasing by means of the SAXS intensities. The light grey curves are obtained from the unbiased simulations of the model interacting with $U_{\alpha}$, the dark grey come from the target model interacting with $U_\text{G\=o}$ and the red lines from the biased simulations.}
\label{fig_ort_protG_saxs}
\end{figure}

\begin{figure}[htp]
\centering
\includegraphics[width=0.95\textwidth]{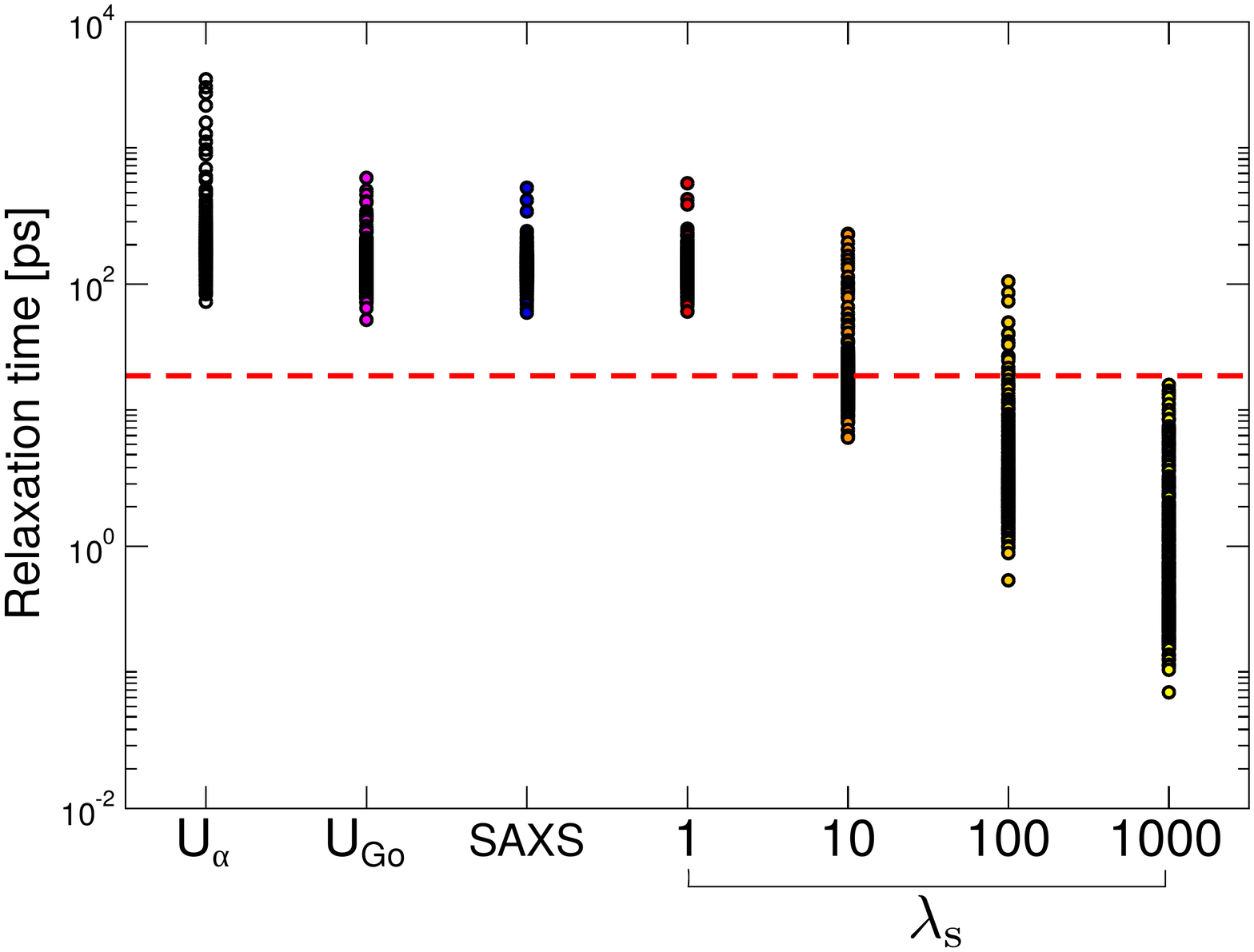}
\caption{Relaxation times for all the variables obtained by TICA analysis on the C$_{\alpha}$ positions. The original unbiased potential $U_{\alpha}$ (white dots) shows a longer relaxation time with respect to the unbiased $U_{\text{G\=o}}$ potential (magenta dots). Without time rescaling, both the caliber-biased simulations shows a good agreement in relaxation time (blue dots for the SAXS-biased one and red dot for the $Q$-biased one) with the target potential. Varying the scaling parameter $\lambda_{s}$, we obtain, as expected, a decrease in relaxation times, which for $\lambda_{s}=100,1000$ becomes comparable to the typical diffusion time of the system (baseline in the plot, calculated as $\tau=l^2\gamma/(k_BT)$, where $l$ is the end to end distance of the intial unfolded conformation and is set to 5 nm, $\gamma$ is the thermostat coupling constant and is set to 1 ps$^{-1}$ and $k_BT$ is the energy unit set to 881.3 kJ/mol ($T=106$ K).}
\label{img:fig_ort_protG_saxs_fluct}
\end{figure}

\end{document}